\documentclass{article}
\usepackage{RR}
\RRNo{6424}
\usepackage{hyperref}
\usepackage{times}
\usepackage{epsf}
\usepackage{graphics}
\usepackage{latexsym}
\usepackage{subfigure}
\usepackage{amssymb,amsmath,amsfonts,amsthm}

\input{psfig.sty}

\RRdate{\today} 
\RRauthor{Daniel Gaff\'e \thanks{I3S Laboratory and CNRS}
  \and Annie Ressouche \thanks{INRIA Sophia Antipolis}
  \and Val\'erie Roy \thanks{CMA ENM Sophia Antipolis}
  \thanks {thanks to S. Moisan and J.P Rigault for their careful reading and their 
           fruitful suggestions}
}

\authorhead{Gaff\'e \& Ressouche \& Roy} 

\RRtitle{Compilation modulaire d'un langage synchrone}

\RRetitle{Modular Compilation of  a Synchronous Language}

\titlehead{LE Synchronous Language}

\RRresume{
Dans ce rapport, nous \'etudions le d\'eveloppement  de syst\`emes critiques.
Les m\'ethodes formelles se sont av\'er\'ees
un moyen efficace pour augmenter la fiabilit\'e de tels  syst\`emes, en particulier
ceux qui requi\`erent une certaine s\'ecurit\'e de fonctionnement.
Neanmoins, le d\'eveloppement d'outils automatiques de sp\'ecification et de v\'erification
est limit\'e entre autre par la taille des mod\`eles formels des syst\`emes ou par
des probl\`emes d'ind\'ecidabilit\'e.
Dans ce travail, nous d\'efinissons un langage r\'eactif synchrone ({\sc le})
d\'edi\'e \`a la sp\'ecification de syst\`emes critiques. Ce faisant, nous  b\'en\'eficions 
d'un cadre formel sur lequel nous nous appuyons pour compiler s\'eparement et valider 
les applications. Plus pr\'ecisement, nous d\'efinissons deux s\'emantiques pour notre
langage: une \textit{s\'emantique comportementale} qui associe \`a un programme l'ensemble de ses
comportements et \'evite ainsi toute ambiguit\'e dans l'interpretation des programmes.
Nous d\'efinissons aussi une \textit{s\'emantique \'equationnelle} dirigeant la compilation de
programmes vers diff\'erentes cibles (code c, code vhdl, synth\'etiseurs fpga, observateurs),
permettant ainsi de traiter des applications logicielles et mat\'erielles et aussi de les valider.
Notre approche est pertinente vis \`a vis des deux principales exigences de r\'eelles
applications critiques: la compilation modulaire permet de traiter des applications
cons\'equentes et l'approche formelle permet la validation. On peut constater que 
le domaine des langages synchrones manque encore de m\'ethodes pour compiler les
programmes de fa\c{c}on efficace et modulaire. Bien sur, certaines approches optimisent
les codes produits d'un facteur important, mais aucune d'entre elles n'envisagent une compilation
modulaire.
}

\RRabstract{
Synchronous languages rely on formal methods to facilitate the developement of
applications in an efficient and reusable way. In fact, formal methods
have been advocated as a means of increasing the
reliability of systems, especially those which are safety or business
critical.
It is even more
difficult to develop automatic specification and verification tools
due to limitations such as  state explosion, undecidability, etc...
In this work, we design a new specification model based on a
reactive synchronous approach. We benefit from a formal
framework well suited to perform compilation and formal validation of
systems.  In practice, we design and implement a special purpose
language ({\sc le}) with two semantic: its \textit{behavioral
semantic} helps us to define a program by the set of its behaviors and
avoid ambiguity in programs interpretation; its \textit{
equational semantic} allows the compilation of programs
into software and hardware targets (C code, Vhdl code, Fpga synthesis,
Model checker input format).
Our approach is relevant
with respect to the two main requirements of critical realistic
applications: modular compilation allows us to deal with large
systems, while model-based approach provides us with formal validation.
There is still a lack of efficient and modular compilation means for
synchronous languages. Despite of relevant attempts to optimize generated code,
no approach considers modular compilation. This report tackles this problem by
introducing a compilation technique which relies on the equational semantic
to ensure modularity completed by a new algorithm
to check  causality cycles in the whole
program without checking again the causalty of sub programs.
}

\RRmotcle{
langage synchrone, compilation modulaire , sémantique comportementale
sémantique constructive equationnelle, modularit\'e, compilation s\'epar\'ee.
}

\RRkeyword{
synchronous language, modular compilation, behavioral semantic,
equational constructive semantic, modularity, separate compilation.
}

\RRprojet{Pulsar}

\RRtheme{\THCog}
\URSophia 

\makeatletter
\newcommand{\xLeftarrow}[2][]{%
    \ext@arrow 0055{\Leftarrowfill@}{#1}{#2}%
}
\def\Leftarrowfill@{\arrowfill@\Leftarrow\Relbar\Relbar}

\newcommand{\xRightarrow}[2][]{%
    \ext@arrow 0055{\Rightarrowfill@}{#1}{#2}%
}
\def\Rightarrowfill@{\arrowfill@\Relbar\Relbar\Rightarrow}

\newcommand{\microsteparrow}[2][]{%
    \ext@arrow 0055{\microsteparrowfill@}{#1}{#2}%
}
\def\microsteparrowfill@{\arrowfill@\relbar\relbar\twoheadrightarrow}

\makeatother

\begin{document}

\parindent=0pt

\makeRR 

\newcommand{\Plus}{$\:\boxplus\:$}
\newcommand{\Mult}{$\:\boxdot\:$}
\newcommand{\Larrow}[1]{\rule[.65mm]{#1}{0.15mm}\hspace*{-1.2mm}\longrightarrow}
\newcommand{\fixed}[1]{\texttt{\small{#1}}}
\newcommand{\lstrl}{{\sc le\ }}
\newcommand{\ple}{{\sc ple\ }}
\newcommand{\restr}[1]{\negthickspace\upharpoonright_{#1}}
\newcommand{\eqsem}[2]{{\left\langle{{#1}}\right\rangle}_{{#2}}}

\def\condmath#1{\leavevmode\ifmmode #1 \else $#1$ \fi}
\newcommand{\Bool}{\condmath{\mbox{I}  \! \mbox{B}}}

\clubpenalty=9999
\widowpenalty=9999

\tableofcontents
\newpage

\section{Introduction}

We address the design of safety-critical control-dominated
systems. By design we mean all the work that must be done from the
initial specification of a system to the embedding of the validated
software into its target site.
The way control-dominated systems work is 
\textit{reactive} in the sense of  D. Harel and
A. Pnueli definition\cite{HP85}: they react to external stimuli at a
speed defined and controlled by the system's environment. The
evolution of a reactive system is a sequence of reactions raised by
the environment. A control-dominated application can then be naturally
decomposed into a set of communicating reactive sub-systems each dealing
with some specific part of the global behavior, combined together to
achieve the global goal.

It is now  stated that general purpose programming languages
are not suited to design reactive systems: they are clearly
inefficient to deal with the inherent complexity of such systems. From
now on, the right manner to proceed is to design languages dedicated
to reactive systems.  To this aim, \textit{synchronous languages} such
as Esterel\cite{berry} and SyncCharts \cite{SyncChart}, 
dedicated to specify event-driven applications; Lustre and
Signal\cite{HAL93}, data flow languages well suited to describe
signal processing applications like, have been designed.
They are model-based languages to allow formal verification of the system
behavior and they agree on three main features:
\begin{enumerate}
\item \textit{Concurrency}: they support functional concurrency and they
rely on notations that express concurrency in a user-friendly manner.
\lstrl adopts an imperative Esterel-like style to express parallelism.
However, the semantic of concurrency is the same for all synchronous languages
and simultaneity of events is primitive.
\item \textit{Simplicity}: the language formal models are simple (usually
mealy machines or netlists) and thus formal reasoning is made tractable.
In particular, the semantic for parallel composition is clean.
\item \textit{Synchrony}: they support a very simple execution model.
First, memory is initialized and then, for each input event set, outputs are
computed and then memory is updated.
Moreover, all mentioned actions are assumed to take finite memory and time.
\end{enumerate}

Synchronous languages rely on the \textit{synchronous hypothesis}
which assumes a discrete logic time scale, made of instants
corresponding to reactions of the system. All the events concerned by
a reaction are simultaneous: input events as well as triggered
output events.  As a consequence, a reaction is instantaneous (we
consider that a reaction takes no time), there are no concurrent
partial reactions, and  determinism is thus ensured.

There are numerous advantages to the synchronous approach. The main
one is that  temporal semantic is simplified, thanks to the
afore mentioned logical time. This leads to clear temporal constructs
and easier time reasoning.
Another key advantage is the reduction of state-space explosion,
thanks again to  discrete logical time: systems evolve in a sequence
of discrete steps, and nothing occurs between two successive steps.
A first consequence is that program debugging, testing, and validating
is  easier. In particular, formal verification of synchronous
programs is possible with techniques like  model checking.
Another consequence is that synchronous language compilers are able to
generate automatically embeddable code, with performances that can be
measured precisely.

Although synchronous languages have begun to face the state explosion
problem, there is still a need for further research on their efficient and
modular compilation.
The initial compilers translated the program into an extended finite
state machine. The drawback of this approach is the potential state
explosion problem. Polynomial compilation was first achieved by a
translation to equation systems that symbolically encode the
automata. This approach is successfully used for hardware synthesis
and is the core of commercial tools \cite{esterelstudio} but the
generated software may be very slow.
Then several approaches translate the program into event graphs
\cite{saxo-rt} or concurrent data flow graphs \cite{cec,jacky} to
generate efficient C code.  All these methods have been used to
optimize the compilation times as well as the size and the execution
of the generated code.

However none of these approaches consider a modular compilation. Some
attempts allow a distributed compilation of programs \cite{saxo-rt,
cec} but no compilation mechanism relies on a modular semantic of
programs. Of course there is a fundamental contradiction in relying on a
formal semantic to compile reactive systems because a perfect
semantic would combine three important properties:
\textit{responsiveness}, \textit{modularity} and
\textit{causality}. Responsiveness means that we can deal with a
logical time and we can consider that output events occur in the same
reaction as the input events causing them.  It is one of the
foundations of the synchronous hypothesis.
Causality means that for each event generated
in a reaction, there is a causal chain of events leading to this
generation; no causality loop may occur.  A semantic is modular when
``environment to component'' and ``component to component'' communication are
treated symmetrically. In particular, the semantic of the composition
of two reactive systems can be deduced from the respective semantic of
each sub-part. Another aspect of modularity is the coherent view each
subsystem has of what is going on. When an event is present, it is
broadcasted all around the system and is immediately available for
every part which listens to it.  Unfortunately, there exists a theorem
(``the RMC barrier theorem'') \cite{causality-modularity} that states
that these three properties cannot hold together in a semantic.
Synchronous semantic are responsive and modular. But causality
remains a problem in these semantic and modular compilation must be
completed by a global causality checking.

In this paper we introduce a reactive synchronous language, we define
its behavioral semantic that gives a meaning to programs and an
equational semantic allowing first, a modular compilation 
and, second, a separate verification of properties.  Similarly to other
synchronous semantic, we must check that programs have no potential causality loop.
As already mentioned, causality can only be checked globally since a bad causality
may be created when performing the parallel composition of two causal sub programs. 
We compile \lstrl programs into equation systems and the program is causal if 
its compilation is cycle free. The major contribution of our approach relies on
the introduction of a new sorting algorithm that allows us to start
from already compiled and checked subprograms to compile and check the overall program
without sorting again all the equations.

\section {LE Language}
\label{lelanguage}

\lstrl language belongs to the family of reactive synchronous
languages.  It is a discrete control dominated language. We first
describe its syntax (the overall grammar is detailed in appendix \ref{legrammar}).

The \lstrl language unit are  \textit{named modules}.
The language's operators and constructions are chosen to fit the
description of reactive applications as a set of concurrent
communicating sub-systems. Communication takes place between modules
or between a module and its environment.
Sub-system communicates via \textit{events}.

The \textit{module interface} declares the set of \textit{input events}
it reacts to and the set of \textit{output events} it emits.
For instance, the following piece of code shows the declarative part
of a \textit{Control} module used in the example in
section \ref{leexample}.

{\small
\begin{verbatim}
module Control:
Input:forward, backward, upward,  downward, StartCycle;
Output:MoveFor, MoveBack, MoveDown, SuckUp, EndCycle ;
\end{verbatim}
}

\subsection{LE Statements}

The \textit{module body} is expressed using a set of \textit{control operators}.
They are the cornerstone of the language
because they operate over event's status. Some operators terminate
instantaneously, some other takes at least one instant. We mainly distinguish
two kinds of operators: usual programming language operators and operators devoted to
deal with logical time.

\subsubsection{Non Temporal Statements}

\lstrl language offers two basic instructions:
\begin{itemize}
\item
The \textit{nothing instruction} does "nothing" and terminates instantaneously.
\item
The \textit{event emission instruction} (\textit{emit speed}) sets to present the
status of the emitted signal.
\end{itemize}

Moreover, some operators help us to built composite instructions:

\begin{itemize}

\item
The \textit{present-then-else instruction} 
(\textit{present S \{ P1\} else \{ P2\}}) is a usual conditional statement except that
bool\-ean combinations of signals status are used as conditions.

\item
In the \textit{sequence instruction} ($P_1 \gg P_2$) 
the first sub-instruction $P_1$ is executed. Then, if $P_1$ terminates
instantaneously, the sequence executes immediately its second
instruction $P_2$ and stops whenever $P_2$ stops. If $P_1$ stops, the
sequence stops. The sequence terminates  at the same instant as its second
sub-instruction $P_2$ terminates. If the two sub-instructions are
instantaneous, the sequence terminates instantaneously.

\item
The \textit{parallel instruction}($P_1 \| P_2$) begins the execution of
its two sub-instructions at the same instant. It terminates when 
both sub-instructions terminate. When the two sub-instructions are
instantaneous, the parallel is instantaneous. Notice that the parallel instruction
agrees with the synchronous hypothesis and allows the simultaneity of trigger signals
causing $P_1$ or $P_2$.

\item
A \textit{strong or weak preemption instruction} over a signal $S$
can surround an instruction $P$  as in: $abort\ P\ when\ S$.
While the signal status evaluates to ``absent'', instruction $P$
continues its execution. The instant the event evaluates to ``present'',
the instruction is forced to terminate. When the instruction is
preempted, the \texttt{weak} preemption let the instruction ends its
current execution while the \texttt{strong} one does not.  If
the instruction terminates normally without been preempted, the
preemption instruction also terminates and the program execution
continues.

\item
A \textit{Loop instruction} ($loop\ \{P\}$) surrounds an instruction $P$. 
Instruction $P$ is automatically restarted the same instant it
terminates. The body of a loop cannot be instantaneous since it will start again the execution
of its body within the same instant. 

\item
\textit{Local signals instruction} ($local\ S\ \{ P \}$) is used to
encapsulate communication channels between two sub systems. The scope
of $S$ is restricted to $P$. As a consequence, each local signal tested
within the body of the local instruction must be emitted from the body.

\item
A \textit{module call instruction}($Run$) is used to run an external module
inside another module. Recursive calls of module are not allowed. 
Running a module does not terminate instantaneously.  In the
declarative part of the module,
you can specify the paths where the already compiled code of the called modules are:

{
\small
\begin{verbatim}
Run: "./TEST/control/" : Temporisation;
Run: "./TEST/control/" : NormalCycle;
\end{verbatim}
}

\end{itemize}

\subsubsection{Temporal Statements}

There are two temporal operators in \lstrl.

\begin{itemize}
\item
The \textit{pause instruction} stops for exactly one reaction.
\item
The \textit{waiting instruction} (\textit{wait S}) waits the presence of a signal.
The first time the execution of the program
reaches a \texttt{wait} instruction, the execution stops (whatever the
signal status is). At the beginning of the following instant, if the
signal status is tested ``present'' the instruction terminates and the
program continues its execution, otherwise it stays stopped. 
\end{itemize}

\subsubsection{Automata Specification}

Because it remains difficult to design an automaton-like behavior
using the previously mentioned operators, our language offers an
\textit{automaton description} as a native construction. 
An automata is a set of states and labeled transitions between states.
Some transitions are initial and start the automata run while terminal states indicate
that the automaton computation is over. The label of transitions
have two fields: a trigger  which is a boolean combination of signal status
and an output  which is the list of signals emitted when the transition is
taken (i.e when the trigger part is true). \lstrl automata are Mealy machines and they
have a set of input signals to define transition triggers and a set of output signals that
can be emitted when a transition is raised.
In \lstrl, the body of a module is either an instruction or an automaton.
It is not allowed to build new instructions by combining instructions and
automata. For instance, the only way to put in parallel an automaton and
the emission of a signal is to call the module the body of which is the automata through
a run operation.
Practically, we offer a syntactic means to describe an automaton (see appendix \ref{legrammar}
for a detailed syntax).
Moreover, our
graphical tool ({\sc galaxy}) helps  users edit automata and
generate the \lstrl code.

\section{LE Behavioral Semantic}
\label{lebehavioral}

\lstrl behavioral semantic is useful to give a meaning to each program and thus
to define its behavior without ambiguity.
To define the behavioral semantic of {\sc le}, we first introduce a logical context
to represent events, then we  define the \lstrl process calculus in order to
describe the behavioral semantic rules.

\subsection{Mathematical Context}

Similarly to others synchronous reactive languages, \lstrl handles
\textit{broadcasted signals} as communicating means.
A program reacts to input events by producing
output events.  An \textit{event} is a signal carrying some
information related to its \textit{status}.  The set of signal status
$\xi$ ($\xi$ = $\{\bot, 0, 1, \top\}$) 
{\footnote{
we also denote true and false values of $\xi$ boolean algebra by $1$ and $0$ by misuse of
language. Nevertheless, when some ambiguity could occur, we will denote them
$1_{\xi},\ 0_{\xi}$.
}}
is intented to record the
status of a signal at a given instant.  Let $S$ be a signal, $S^x$
denotes its instant current status.  More precisely, $S^1$
means that $S$ is present, $S^0$ means that $S$ is
absent, $S^{\bot}$ means that $S$ is neither present
nor absent  and finally $S^{\top}$ corresponds to
an event whose status cannot be induced because it has two
incompatible status in two different sub parts of the program.
For instance, if $S$ is both absent and present,
then it turns out to have $\top$ status and thus an error
occurs. 
Indeed. the set $\xi$ is a complete lattice with the $\leq$ order:
\renewcommand\arraystretch{0.5}
\[
\begin{array} {c c c}
&\top&\\
\nearrow & & \nwarrow\\
0\quad  & \uparrow  &\quad  1 \\
\nwarrow& &\nearrow\\
&\bot&\\
\end{array}
\]
\renewcommand\arraystretch{1.0}

\subsubsection*{Composition Laws for $\xi$}

We define 3 internal composition laws in $\xi$: {\Plus}, \Mult and $\neg$ (to extend
the usual operations defined for classical boolean set $\Bool$), as follows:

The {\Plus} law is a binary operation whose result is the upper bound of its
operands:

\begin{center}
\begin{tabular}{| c || c | c | c | c |}
\hline
{\Plus}  & $1$     & $0$   & $\top$ & $\bot$\\
\hline\hline
$1$     & $1$ & $\top$  & $\top$ & $1$\\
\hline
$0$     & $\top$  & $0$  & $\top$ & $0$\\
\hline
$\top$  & $\top$  & $\top$  & $\top$ & $\top$\\
\hline
$\bot$  & $1$ & $0$  & $\top$ & $\bot$\\
\hline
\end{tabular}
\end{center}

Particularly:
\begin{itemize}
\item  $\bot $ \Plus $ \bot = \bot$;
\item $1 $\Plus $0  = 0 $\Plus$ 1 = \top$;
\item $\top$ is an absorbing element;
\end{itemize}

The \Mult law is a binary operation whose result is the lower bound of its
operands:

\begin{center}
\begin{tabular}{| c || c | c | c | c |}
\hline
\Mult  & $1$     & $0$   & $\top$ & $\bot$\\
\hline\hline
$1$     & $1$ & $\bot$  & $1$ & $\bot$\\
\hline
$0$     & $\bot$  & $0$  & $0$ & $\bot$\\
\hline
$\top$  & $1$  & $0$  & $\top$ & $\bot$\\
\hline
$\bot$  & $\bot$ & $\bot$  & $\bot$ & $\bot$\\
\hline
\end{tabular}
\end{center}

Particularly:
\begin{itemize}
\item  $\top \:\boxdot\: \top = \top$;
\item $1\:\boxdot\:  0  = 0 \:\boxdot\: 1 = \bot$;
\item $\bot$ is an absorbing element;
\end{itemize}

Finally, the {$\neg$} law is an inverse law in $\xi$:

\begin{center}
\begin{tabular}{| c | c |}
\hline
{$x$} & $\neg\ x$\\
\hline
$1$     & $0$\\
\hline
$0$      & $1$\\
\hline
$\top$      & $\bot$\\
\hline
$\bot$  & $\top$\\
\hline
\end{tabular}
\end{center}

The set $\xi$ with these 3 operations verifies the axioms of
\textit{Boolean Algebra}: commutative and associative
axioms for \Plus and \Mult,
distributive axioms both for \Mult over \Plus and for \Plus over \Mult,
neutral elements for \Plus and \Mult and complementarity.

\begin{center}
\begin{tabular}{|l |c| c r|}
\hline
Commutativity: & $x$ \Plus $y = y$ \Plus $x$ & $x \:\boxdot\: y = y \:\boxdot\: x$ & (1) \\
\hline
Associativity: &$(x$ \Plus $y)$ \Plus $x = x $ \Plus $ (y $\Plus $z)$ & 
                      $(x \:\boxdot\: y) \:\boxdot\: x = x \:\boxdot\:t (y \:\boxdot\: z)$ & (2) \\
\hline
Distributivity: & $x \:\boxdot\: (y $\Plus$ z) = (x \:\boxdot\: y)$ \Plus $ (x \:\boxdot\: z)$ &
                   $x $\Plus $(y \:\boxdot\: z) = (x $\Plus $y) \:\boxdot\: (x $\Plus $z)$  & (3)\\
\hline
Neutral elements: & $x $\Plus $\bot = x$ & $x \:\boxdot\: \top = x$ & (4)\\
\hline
Complementarity: & $x $\Plus $\neg\ x = \top $ & $x \:\boxdot\: \neg\ x = \bot$ & (5)\\
\hline
\end{tabular}
\end{center}

Axioms (1) and (4) are obvious looking at the previous tables that define the
\Plus and \Mult laws. Axioms (2) and (4) are also obviously true but their proofs necessitate to
compute the appropriate tables. Finally, axiom (5) results from the following table:

\begin{center}
\begin{tabular}{|c|| c | c |}
\hline
$x$  & $x$ \Plus $\neg\ x$ & $x \:\boxdot\: \neg\ x$\\
\hline\hline
$1$     & $1$ \Plus $0 = \top$ & $1 \:\boxdot\: 0 = \bot$ \\
\hline
$0$     &  $0$ \Plus$ 1 = \top$ & $0 \:\boxdot\: 1 = \bot$ \\ 
\hline
$\top$  & $\top$ \Plus $\bot = \top$ & $\top \:\boxdot\: \bot = \bot$\\    
\hline
$\bot$  & $ \bot$ \Plus $\top = \top$ & $ \bot \:\boxdot\: \top = \bot$\\
\hline
\end{tabular}
\end{center}

As a consequence, {\bf $\xi$ is a Boolean algebra} and the following theorems are
valid:

\begin{center}
\begin{tabular}{|c| c| c|}
\hline
Identity law: & $x$ \Plus $x = x$ & $x \:\boxdot\: x = x$ \\
\hline
Redundancy law: & $x$ \Plus $(x \:\boxdot\: y) = x$ & $ x \:\boxdot\: (x $\Plus $y) = x$ \\
\hline
Morgan law: & $\neg\ (x $ \Plus $y) = \neg\ x \:\boxdot\: \neg\ y $&
                     $\neg\ (x \:\boxdot\: y) = \neg\ x $\Plus $\neg\ y $\\
\hline
Neutral element: & $x$ \Plus $\top = \top $& $x \:\boxdot\: \bot = \bot$\\
\hline
\end{tabular}
\end{center}

In such a setting, $xor$, $nor$, $nand$, $\Leftrightarrow$, $\Rightarrow$ are defined:

\begin{tabular}{l c c}
$x\ xor\ y$  & = & $x \:\boxdot\: \neg\ y $\Plus $y \:\boxdot\: \neg\ x $\\
$x\ nor\ y$ & = & $\neg\ x \:\boxdot\: \neg\ y $\\
$x\ nand\ y$ & = & $\neg\ x $\Plus $\neg\ y$\\
$x \Leftrightarrow y $ & = & $(\neg\ x \:\boxdot\: \neg\ y)$ \Plus $(x \:\boxdot\: y)$ \\
$x \Rightarrow y$ & = & $\neg x$ \Plus $y$ \\
\end{tabular}

Hence, we can apply these classical results concerning
Boolean algebras to solve equation systems whose variables belong to
$\xi$. For instance, the equational semantic detailed in section~\ref{leequational}
relies on boolean algebra properties to compute signal status as solution
of status equations.

Moreover, since $\xi$ is a lattice, the \Plus and
\Mult operations are monotonic:
let $x$, $y$ and $z$ be elements of $\xi$, 
($x \leq y) \Rightarrow (x\: \boxplus\: z \leq y\: \boxplus\: z$) and
($x \leq y) \Rightarrow (x\: \boxdot\: z \leq y\: \boxdot\: z)$.

\subsubsection*{Condition Law}
\label{lecondlaw}

We introduce a \textit{condition} law ($\blacktriangleleft$) in $\xi$ to drive a signal status
with a boolean condition:

\begin{center}
\begin{tabular}{c c c}
$\xi \ \times \ \Bool$ & $ \longrightarrow$ & $\xi$\\
$(x,c)$ & $\longmapsto $ & $ x \blacktriangleleft c$\\
\end{tabular}
\end{center}

This law is defined by the following table:

\begin{center}
\begin{tabular}{| c | c | c |}
\hline
{$x$} & $c$ & $x \blacktriangleleft c$\\
\hline
$1$     & $0$ & $\bot$\\
$0$      & $0$ & $\bot$\\
$\top$      & $0$ & $\bot$\\
$\bot$     & $0$ & $\bot$\\
\hline
$1$     & $1$ & $1$\\
$0$      & $1$ & $0$\\
$\top$      & $1$ & $\top$\\
$\bot$     & $1$ & $\bot$\\
\hline
\end{tabular}
\end{center} 

This condition law allows us to change the status of an event according to  a boolean 
condition. It will be useful to define both \lstrl  behavioral and equational semantic
since the status of signals depend of the termination of the instructions that compose a
module. Intuitively, a signal keeps its status if the condition is true, otherwise 
its status is set to $\bot$.

\subsubsection*{Relation between $\xi$ and $\Bool^2$}
\label{bool-encoding}

$\xi$ is bijective to $\Bool \times \Bool$. We define the following encoding:

\begin{center}
\begin{tabular}{| c | c |}
\hline
signal status & encoding \\
\hline
$1$ & $11$\\
$0$ &  $10$\\
$\top$ &  $01$\\
$\bot$ & $00$\\
\hline
\end{tabular}
\end{center}
Hence, a signal status is encoded by 2 boolean variables. 
The first boolean variable of  the status of a signal ($S$) is called its 
definition ($S_{def}$), while the second one is called its value ($S_{val}$).
According to the encoding law, when $S_{def} = 0$ the signal $S$ has either $\top$ or $\bot$ value
for status and it is not defined as present or absent. On the opposite, when
$S_{def} = 1$, the signal is  either present or absent. It is why we choose to
denote the first boolean projection of a signal status by $S_{def}$.

\Bool is the classical boolean set with 3 operators \textit{and} (denoted .), \textit{or}
(denoted +) and \textit{not} (denoted $\overline{x}$, for boolean $x$). 
According to the previous encoding of $\xi$ into $\Bool \times \Bool$ and after algebraic
simplification, we have the following equalities related to \Plus, \Mult and
$\neg$ operators. Let $X$ and $Y$ be 2 elements of $\xi$:

\begin{center}
\begin{tabular} {l c l}
($X$ \Plus $Y)_{def}$ & = & $X_{def}.\overline{Y_{def}}.\overline{Y_{val}} +
                             Y_{def}.\overline{X_{def}}.\overline{X_{val}} +
                             (X_{def}.Y_{def}).\overline{(X_{val} \oplus Y_{val})}$ \\

($X$ \Plus $Y)_{val}$ & = & $X_{val} + Y_{val}$\\
\\
($X \:\boxdot\: Y)_{def}$ & = & $X_{def}.\overline{Y_{def}}.Y_{val} +
                             Y_{def}.\overline{X_{def}}.X_{val} +
                             (X_{def}.Y_{def}).\overline{(X_{val} \oplus Y_{val})}$ \\

($X \:\boxdot\: Y)_{val}$ & = & $X_{val} . Y_{val}$\\
\\
($\neg\ X)_{def}$ & = & $X_{def}$\\
($\neg\ X)_{val}$ & = & $\overline{X_{val}}$\\
\\
$(X \blacktriangleleft c)_{def}$ & = & $X_{def}.c$\\
$(X \blacktriangleleft c)_{val}$ & = & $X_{val}.c$\\
\end{tabular}
\end{center}
where $\oplus$ is the \textit{exclusive or} operator of classical boolean set.
The proof of the last equality is detailed in appendix \ref{appendix-lecondlaw}.

On the opposite side, we can expand each boolean element into a status
member, 0 correspond to 0, and 1 to 1. More precisely let $x$ be an
element of \Bool and $\xi(x)$ its corresponding status, then
$\xi(x)_{def} = 1$ and $\xi(x)_{val} = x$.

\subsubsection*{Notion of Environment}

An \textit{environment} is a finite set of events. Environments are useful to
record the current status of signals  in a reaction. Thus a signal has a unique status
in an environment: if $S^x$ and $s^y$ belongs to the same environment, then $x = y$.

We extend the operation defined in $\xi$ to environments. Let $E$
and $E '$ be 2 environments:

\begin{center}
\begin{tabular}{l c l}
$E$ \Plus  $E '$ & = & $\{S^z | \exists S^x \in E, S^y \in E ' , z = x$\Plus $y\}$\\
$E \:\boxdot\: E '$ & = & $\{S^z | \exists S^x \in E, S^y \in E ', z = x \:\boxdot\: y\}$\\
$\neg E$ & = & $ \{S^{\neg\ x} | \exists S^x \in E\}$\\
$ E\ \blacktriangleleft\ c$ & = & $\{S\ \blacktriangleleft\ c\ |\  S \in E\}$
\end{tabular}
\end{center}

We define a relation ($\preceq$) on environments as follows:
\[
E \preceq E '\ {\rm iff}\ \forall S^x \in E, \exists S^y \in E ' | S^x \leq S^y
\]

Thus $E \preceq E '$ means that $E$ is included in $E '$ and that each element of
$E$ is less than an element of $E '$ according to the lattice order of $\xi$.
As a consequence, the $\preceq$ relation is a total order on environments and 
\Plus and \Mult operations are monotonic according to $\preceq$.

Finally, we will denote $E^{\top}$, the environment where all events have $\top$ status.

\subsection{LE Behavioral Semantic}

\renewcommand\arraystretch{0.5}

In order to describe the behavioral semantic of \lstrl, we first
introduce a process algebra associated with the language. Then we can
define the semantic with a set of rewriting rules that determines a
program execution.  The semantic formalize a reaction of a program
$P$ according to an event input set.  
$P \begin{array}{c} E ' \\ \boldsymbol{\longmapsto} \\ E \end{array} P'$ 
has the usual meaning: $E$ and $E
'$ are respectively input and output environments; program $P$ reacts
to $E$, reaches a new state represented by $P'$ and the output
environment is $E '$. To compute such a reaction we rely on the
behavioral semantic of \lstrl. This semantic supports a rule-based
specification to describe the behavior of each operator of \lstrl
process algebra associated with \lstrl language.  A rule 
has the form: $p \xrightarrow[E]{E ', TERM} p'$ where $p$ and
$p'$ are elements of \lstrl process algebra. $E$ is an environment
that specifies the status of the signals declared in the scope of $p$,
$E '$ is the output environment and $TERM$ is a boolean flag true when
$p$ terminates. This notion of termination differs from the
one used in Esterel language successive behavioral semantic. It means
from the current reaction, $p$ is able to terminate and this information will be sustained until 
the real termination occurs.

Let $P$ be a \lstrl program  and $p$ its corresponding process algebra term.
Given an input event set $E$, a reaction is computed as follows:
\[
P \begin{array}{c}  E ' \\ \boldsymbol{\longmapsto} \\ E \end{array} P'\hspace*{0.5cm} {\rm iff}
\hspace*{0.5cm} p \xrightarrow[E]{E ',\ TERM} p'
\]

\renewcommand\arraystretch{1.0}

\subsubsection* {LE Process Calculus (PLE)}

The \ple process algebra associated to \lstrl language is defined as follows:

\begin{itemize}
\item \fixed{nothing}; 
\item \fixed{halt};
\item !\fixed{s} (emit \fixed{s});
\item \fixed{wait s};
\item \fixed{iwait s} (wait immediate \fixed{s});
\item \fixed{s} ? $p$ : $q$ (present \fixed{s} \{$p$\} else \{$q$\});
\item $p \| q$; 
\item $p \gg q$; 
\item $p \uparrow_{\fixed{s}}$ (abort \{$p$\} when \fixed{s});  
\item $p*$ (loop \{p\});
\item $p \backslash \fixed{s}$ (local \fixed{s} \{$p$\});
\item ${\cal A}({\cal M}, {\cal T}, {\cal C}ond, M_f, {\cal O}, \lambda)$.
Automata ${\cal A}$ is a structure made of 6 components:
\begin{enumerate}
\item  a finite set of {\em macro states} (${\cal M}$).
Each macro state $M$ may be is itself composed of a sub term $p$ (denoted $M[p]$);
\item a finite set of {\em conditions} (${\cal C}ond$);
\item a finite set of {\em transitions} (${\cal T}$). 
A transition is a 3-uple $<M, c, M'>$ where  $c \in {\cal C}ond$ is a boolean
condition raising the transition from macro state $M$ to macro state $M'$.
We will denote $M \rightarrow M'$ for short in the rest of the report and $c_{M \rightarrow M'}$ 
will denote the  condition associatesd with the transition.
.
${\cal T}$ is also composed of initial transitions of the form:
${\rightarrow M'}$. They are useful to start the automata run. When condition
$c$ is true, the macro state $M'$ is reached;
\item a {\em final} macro state $M_f$;
\item a finite set of {\em output signals} (${\cal O}$) paired with an {\em output function} $\lambda$ that
links macro states and output signals:
$\lambda: {\cal T} \longrightarrow {\cal P}({\cal O})$, defined as follows:
$\lambda({M \rightarrow M'}) = o \subseteq {\cal O}$ is the set of output signals emitted when
the trigger condition $c_{M \rightarrow M'}$ is true.
\end{enumerate}
\end{itemize}

Each instruction of \lstrl has a natural translation as an operator of the process
algebra. As a consequence, we associate a  term of the process algebra with the body of each
program while the interface part allows  to build the global environment useful to
define the program reaction as a rewriting of the behavioral semantic.
Notice that the operator \fixed{iwait s} does not correspond to any instruction of the
language, it is introduced to express the semantic of the \fixed{wait} statement.
It is a means to express that the behavior of a term takes at least one instant.
It is the case of \fixed{wait s} that skip an instant before reacting to the presence of
\fixed{s}.

More precisely, we introduce a mapping: $\Gamma$ : \lstrl $\rightarrow$ \ple, which associates 
a \ple term with each \lstrl program. $\Gamma$ is defined according to the syntax of the
\lstrl language.

Let $P$ be a \lstrl program, $\Gamma(P)$ is structurally defined on the body of $P$.
\begin{itemize}
\item $\Gamma({\rm nothing}) = \fixed{nothing}$;
\item $\Gamma({\rm halt}) = \fixed{halt}$;
\item $\Gamma({\rm emit}\ \fixed{s}) = !\fixed{s}$;
\item $\Gamma({\rm wait}\ \fixed{s}) = \fixed{wait s}$;
\item $\Gamma({\rm present} \fixed{s} P_1\ {\rm else}\ P_2) = 
\fixed{s} ? \Gamma(P_1) : \Gamma(P_2)$;
\item $\Gamma(P_1 \| P_2) = \Gamma(P_1) \| \Gamma(P_2)$;
\item $\Gamma(P_1 \gg P_2) = \Gamma(P_1) \gg \Gamma(P_2)$;
\item $\Gamma({\rm abort}\ P_1\ {\rm when}\ \fixed{s}) = \Gamma(P_1) \uparrow_{\fixed{s}}$;
\item $\Gamma({\rm loop}\ \{P_1\}) = \Gamma(P_1)*$;
\item $\Gamma({\rm local}\ \fixed{s} \{P_1\}) = \Gamma(P_1) \backslash \fixed{s}$;
\item $\Gamma({\rm run}\ P_1) = $ wait \texttt{tick} $\gg \Gamma(P_1)$ where
\texttt{tick} is a ``clock'' signal present in each reaction; 
\item $\Gamma({\cal A}({\cal M}, {\cal T}, {\cal C}ond, M_f, {\cal O}, \lambda)$ =
${\cal A}({\cal M}, {\cal T}, {\cal C}ond, M_f, {\cal O}, \lambda)$.
\end{itemize}

\subsubsection*{Behavioral Semantic Rules}

\renewcommand\arraystretch{2.0}

The basic operators of \lstrl process algebra have the following
rewriting rules.  Both \fixed{nothing} and \fixed{halt} have no
influence on the current environment, but the former is always ready to leave
and the latter never.
The emit operator is ready to leave and the signal emitted is set present in the environment
{\footnote{
In the following, we will denote $\fixed{s} \leftarrow 1$ the setting of \fixed{s}'value to 1 
($\xi(\fixed{s})= 1$)}}. 

\[
\begin{array} {l l l r}
\fixed{nothing}& \xrightarrow[E] {E,\ 1}& \fixed{nothing} &\qquad\qquad (nothing) \\
\fixed{halt} & \xrightarrow[E] {E,\ 0}& \fixed{nothing} & \qquad\qquad (halt)\\
!\fixed{s}& \xrightarrow [E] {E[\fixed{s} \leftarrow 1],\ 1} &
\fixed{nothing} & \qquad\qquad (emit)\\
\end{array}
\]

\subsubsection*{Wait}

The semantic of
\fixed{wait} is to wait at least one instant. Thus, to express its
behavior, we introduce the \fixed{iwait} operator. Then, \fixed{wait
s} is not ready to leave, and rewrites into \fixed{iwait s}. This rewriting
behaves like \fixed{wait s} except that it reacts instantaneously to the
signal presence.

\[
\fixed{wait s}  \xrightarrow[E] {E,\ 0} {\fixed{iwait s}}
\qquad\qquad (wait)
\]
\[
\frac {\fixed{s}^1 \in E}{\fixed{iwait s}\xrightarrow[E] {E,\ 1} \fixed{nothing}}
\quad (iwait1)\qquad 
\frac {\fixed{s}^1 \not\in E}{\fixed{iwait s}\xrightarrow[E] {E,\ 0} \fixed{iwait s}}
\quad (iwait2) \qquad
\]

\subsubsection*{Present}
\label{present-behav}

The semantic of $\fixed{s}\ ?\ p\ :\ q$ operator depends on the status of \fixed{s}
in the initial environment $E$.
If \fixed{s} is present
(resp absent) in $E$, the operator behaves like $p$ (resp $q$)
(rules $present1$ and $present2$).
Otherwise, if \fixed{s} is undefined we cannot progress in the rewriting system (rule $present3$)
and if the computation of \fixed{s} internal status results in $\top$, it is an error and
this last is propagated (each event is set to error in the environment).

\[
\frac { p \xrightarrow[E]{E_p,\ TERM_p} p',  q \xrightarrow[E]{E_q,\ TERM_q} q',
        \fixed{s}^1 \in E}
      {\fixed{s}\ ?\ p\ :\ q \xrightarrow[E]
                               {E_p,\ TERM_p} p'}
\quad(present1)\qquad
\]   

\[
\frac { p \xrightarrow[E]{E_p,\ TERM_p} p',  q \xrightarrow[E]{E_q,\ TERM_q} q',
        \fixed{s}^0 \in E}
      {\fixed{s}\ ?\ p\ :\ q \xrightarrow[E]
                               {E_q,\ TERM_q} q'}
\quad(present2)\qquad
\]   

\[
\frac { p \xrightarrow[E]{E_p,\ TERM_p} p',  q \xrightarrow[E]{E_q,\ TERM_q} q',
        \fixed{s}^{\bot} \in E}
      {\fixed{s}\ ?\ p\ :\ q \xrightarrow[E]{E,\ 0} 
        \fixed{s}\ ?\ p\ :\ q}
\quad(present3)\qquad
\]   

\[
\frac { p \xrightarrow[E]{E_p,\ TERM_p} p',  q \xrightarrow[E]{E_q,\ TERM_q} q',
        \fixed{s}^{\top} \in E}
      {\fixed{s}\ ?\ p\ :\ q \xrightarrow[E]{E^{\top},\ 1} 
        \fixed{s}\ ?\ p\ :\ q}
\quad(present4)\qquad
\]  

\subsubsection*{Parallel}

The parallel operator computes its two arguments according to the broadcast of signals
between both sides and it terminates when both sides do.

\[
\frac{p \xrightarrow[E]{E_p,\ TERM_p}p'\quad ,\quad q \xrightarrow[E]{E_q,\ TERM_q}q'}
     {p \| q \xrightarrow[E]{E_p \:\boxplus\: E_q,\:TERM_p . TERM_q} p'\|q'}
\quad(parallel)\qquad
\]

\subsubsection*{Sequence} 

The sequence operator has the usual behavior. While the first argument does not terninate
we don't begin the computation of the second argument (rule $sequence1$). When it 
terminates , we start the second argument (rule $sequence2$).

\[
\frac {p \xrightarrow[E]{E_p,\ 0} p'}
      {p \gg q \xrightarrow[E]{E_p,\ 0} p' \gg q}
\quad(sequence1)\qquad
\frac { p \xrightarrow[E]{E_p,\ 1}\ \fixed{nothing},  
        q \xrightarrow[E_p]{E_q,\ TERM_q} q'}
      {p \gg q \xrightarrow[E]{E_q ,\ TERM_q} q'}
\quad(sequence2)\qquad
\]

\subsubsection*{Abort}

The behavior of the abort operator first derives the body of the statement.
Thus, if the aborting  signal is present is the input
environment, then the statement rewrites in \fixed{nothing} and terminates (rule $abort1$).
If it is not, the body of the statement is derived again (rules
$abort2$ and $abort3$)

\[
\frac {p \xrightarrow[E]{E_p,TERM_p} p', {\fixed{s}^1 \in E}}
      {p \uparrow_{\fixed{s}} \xrightarrow [E] {E_p, \ 1}\fixed{nothing}} 
\quad(abort1)\qquad
\]

\[
\frac {\ p \xrightarrow[E]{E_p, 1} \fixed{nothing}, \fixed{s}^1 \not\in E}
      {p {\uparrow}_{\fixed{s}} \xrightarrow[E]{E_p, 1} \fixed{nothing}}
\quad (abort2) \qquad
\frac {\ p \xrightarrow[E]{E_p, 0} p', \fixed{s}^1 \not\in E}
      {p {\uparrow}_{\fixed{s}} \xrightarrow[E]{E_p, 0} p' {\uparrow}_{\fixed{s}}} 
\quad (abort3) \qquad
\]

\subsubsection*{Loop}

Loop operator never terminates and $p*$ behaves as 
$p \gg p*$.

\[
\frac {p \xrightarrow[E]{E_p, 0} p'}
      {p* \xrightarrow[E]{E_p,\ 0} p' \gg p*}
\quad(loop)\qquad
\]

\subsubsection*{Local}

Local operator behaves as an encapsulation. Local signals are no longer visible
in the surrounding environment.

\[
\frac {p \xrightarrow[E \cup \fixed{s}^{\bot}] {E_p,\ TERM_p} p'}
      {p \backslash \fixed{s} \xrightarrow [E]{E_p-\{\fixed{s}\}, TERM_p} p' \backslash \fixed{s}}
\quad(local)\qquad
\]

\subsubsection*{Automata}

Automata are deterministic (i.e $\forall M \in {\cal M},\ \exists!
{M \rightarrow M'} \in {\cal T}$ such that $c_{M \rightarrow M'}= 1$).

The semantic of automata terms relies on macro state semantic.  A macro state
does not terminate within a single reaction. Its duration is at least one
instant. Thus, $M[p]$ waits an instant and then has the same behavior than p.

\[
 M [p] \xrightarrow[E]{E,\ 0} p
\]

If the macro state is only a state without sub term $p$, then 
\[
M \xrightarrow[E]{E,\ 0} nothing
\]

Now, we define the rewriting rules for automata ${\cal A}$. 
The evaluation of a condition $c \in {\cal C}ond$ depends on the current status of
signals in the  environment.
To denote the current value of a condition we will use the following notation:
$E \models c = b $

Axiom:

\[
\frac {{\rightarrow M}\in {\cal T}, E \models c_{\rightarrow M} = 1}
 {{\cal A} \xrightarrow[E]{E[s \leftarrow 1 | s \in 
                                             \lambda({{\rightarrow M}})],\ 0} <{\cal A}, M, M[p]>}
\quad(automata0)\qquad
\]

Rewriting rules for automata describe the behavior of a reaction as usual. Thus,
we define rewriting rules on a 3-uple: $<{\cal A}, M, p>$. The first element of the tuple is
the automaton we consider, the second is the macro state we are in, and the third is the
current evaluation of the sub term involved in this macro state.

\[
\frac {p  \xrightarrow [E]{E_p, TERM} p', 
          \quad \forall M'\ {\rm such\ that\ } {M \rightarrow M'} \in {\cal T}\ 
                E_p \models\ \sum c_{M \rightarrow M'} \neq 1 }
      {<{\cal A}, M, p> \xrightarrow [E]{E_p,\ 0} <{\cal A}, M , p'>}
\quad(automata1)\qquad 
\]
\[
\frac{\exists M' \  {\rm such \ that }\  {M \rightarrow M'} \in {\cal T}
              {\rm \ and\ } E \models  c_{M \rightarrow M'} = 1 }
      {<{\cal A}, M, p>  \xrightarrow[E] 
                {E [s \leftarrow 1 | s \in \lambda({{M \rightarrow M'}})] ,\ 0} <{\cal A}, M', M'[p]>}
\quad(automata2)\qquad 
\]
\[
\frac {p  \xrightarrow [E]{E_p, 1} nothing}
      {<{\cal A}, M_f, p> \xrightarrow [E]{E_p,\ 1} nothing}
\quad(automata3)\qquad 
\]

Rule $automata0$ is the axion to start  the evaluation of the automaton.
Rule $automata1$ expresses the behavior of ${\cal A}$ automata  when
all the transition trigger conditions are false: in such a case, the sub term associated with the
current macro state is derived (whatever the derivation is) and the automata does not terminate.
On the opposite side, rule $automata2$ expresses the
automata behavior when a transition condition becomes true. In such a case, the automata
steps to the next macro state specified in the condition and the emitted signals associated
with the transition are set to 1 in the environment.
Finally, rule $automata3$ is  applied when the evaluation of the term included in the 
final macro state is over; then the automata computation is terminated.

\renewcommand\arraystretch{1.0}

The behavioral semantic is a ``macro'' semantic that gives the meaning of  a reaction
for each term of the \lstrl process algebra. Nevertheless, a reaction is the least fixed
point of a micro step semantic that computes the output environment from the initial one.
According to the fact that the \Plus and \Mult operations are monotonic with respect to the
$\preceq$ order, we can rely on the work about  denotational semantic~\cite{denotational}
to ensure that for each term, this least fixed point exists. 
Practically, we have $ p \xrightarrow[E]{E '} p'$ if there is a sequence of micro steps
semantic:
\[
\ p {\microsteparrow [E] {E_1}} p_1,  p_1 {\microsteparrow [E_1] {E_2}} p_2, .....
\]
At each step $E_{i+1} = F^i(E_i)$ , since the $F_i$ functions are some combinations of
\Plus  operator and $\blacktriangleleft$ condition law, they are monotonic and 
then $\forall i , E_{i+1} \preceq F^i(E_i)$.
Then, we have $E ' = \sqcup_{n} F^n(E_n)$, thus it turns out that $E '$ is the least fixpoint 
 of the family of $F^n$ functions. But, $\xi$ boolean algebra is a complete lattice, then
so is the set of environments, as a  consequence  such a least fixpoint exits.

\section{LE Equational Semantic}   
\label{leequational}

In this section, we introduce a constructive circuit semantic
for \lstrl which gives us a practical means to compile \lstrl programs in a modular way.

The behavioral semantic describes how the program reacts in an instant. It is logically
correct  in the sense that it computes a single output environment for each input event
environment when there is no causality cycles.
To face this  causality cycle problem specific to synchronous approach, constructive
semantic have been introduced~\cite{berry-book}. Such a semantic for synchronous languages
are the application of constructive boolean logic theory to synchronous language semantic
definition. The idea of constructive semantic is to ``forbid self-justification and 
any kind of speculative reasoning replacing them by a fact-to-fact propagation''.
In a reaction, signal status are established following propagation laws:
\begin{itemize}
\item each input signal status is determined by the environment;
\item each unknown signal $S$ becomes present if an ``emit $S$'' can be executed;
\item each unknown signal $S$ becomes absent if  an ``emit $S$'' cannot be executed;
\item the then branch of a test is executed if the signal test is present;
\item the then branch of a test is not executed if the signal cannot be present;
\item the else branch of a test is executed if the signal test  is absent;
\item the else branch of a test is not executed if the signal test cannot be absent;
\end{itemize}

A program is \textit{constructive} if and only if fact propagation is
sufficient to establish the presence or absence of all signals. 

An elegant means to define a constructive semantic for a language is to
translate each program into a constructive circuit.
Such a translation ensures that programs containing no cyclic instantaneous 
signal dependencies are translated into cycle free circuits.
Usually, a boolean
sequential circuit is defined by a set of wires $W$, a set of
registers $R$, and a set of boolean equations to assign values to wires and
registers.  $W$ is partitioned into a set of input wires $I$,
output wires $O$ and a set of local wires. The circuit computes output wire
values from input wires  and register values. Registers are boolean memories that feed back
the circuit. The computation of circuit outputs is done according to a propagation
law and to ensure that this propagation leads to logically correct solutions, a constructive
value propagation law is supported by the computation.

\subsubsection*{Constructive Propagation Law}

Let ${\cal C}$ be a circuit, $I$ its input wire set, $R_{v}$  a register valuation (also called
a ``state'') and $w$ a wire expression.  Following \cite{berry-book}, the constructive propagation 
law has the form : $I,R_{v} \vdash w \hookrightarrow b$, $b$ is a boolean value and
the law means that under $I$ and $R$ assumptions, $w$ evaluates to $b$.
The definition of the the law is:

\[
\begin{array} {l c l}
I,R_{v} \vdash b \hookrightarrow b &&\\
I,R_{v} \vdash w \hookrightarrow b &{\rm if}& I(w) = b \\
I,R_{v} \vdash w \hookrightarrow b &{\rm if}& R(w) = b \\
I,R_{v} \vdash w \hookrightarrow b &
                       {\rm if}& w = e \in {\cal C}, I,R_{v} \vdash e \hookrightarrow b \\
I,R_{v} \vdash w \hookrightarrow b &{\rm if}& 
                   w = \overline{e}, I,R_{v} \vdash e \hookrightarrow \overline{b }\\
I,R_{v} \vdash w \hookrightarrow 1 &
                        {\rm if}& w = e + e', I,R_{v} \vdash e \hookrightarrow 1 \ {\rm or}\ 
                        I,R_{v} \vdash e' \hookrightarrow 1\\
I,R_{v} \vdash w \hookrightarrow 0 &{\rm if}& 
                               w = e + e', I,R_{v} \vdash e \hookrightarrow 0 \ {\rm and}\ 
                               I,R_{v} \vdash e' \hookrightarrow 0\\
I,R_{v} \vdash w \hookrightarrow 1 &{\rm if}& 
                             w = e . e', I,R_{v} \vdash e \hookrightarrow 1 \ {\rm and}\ 
                             I,R_{v} \vdash e' \hookrightarrow 1\\
I,R_{v} \vdash w \hookrightarrow 0 &{\rm if}& 
                               w = e . e', I,R_{v} \vdash e \hookrightarrow 0 \ {\rm or}\ 
                               I,R_{v} \vdash e' \hookrightarrow 0\\
\end{array}
\]

The $\hookrightarrow$ propagation law is the logical characterization of constructive circuits.
Nevertheless, this notion also supports two equivalent  characterizations.
The denotational one relies on three-values boolean ($\Bool_{\bot} = \{\bot, 0, 1 \}$)
and a circuit ${\cal C}$ with $n$ wires, input wire set $I$ and registers $R$ is
considered as  a monotonic
function ${\cal C}(I,R): \Bool_{\bot}^n \longrightarrow \Bool_{\bot}^n$. Such a function has a
least fixed point and this latter is equal to the solution of the equation system associated
to the logical point of view.
On the other hand, the electric characterization uses the inertial delay model of 
Brozowski and requires electric stabilization for all delays. In \cite{Shiple},
it is shown that a circuit ${\cal C}$ is constructive for $I$ and $R$ if and only if
for any delay assignment, all wires stabilize after a time t. The resulting electrical
wire values are equal to logical propagation application results.
 
\subsection{Equational Semantic Foundations}

\lstrl circuit semantic associates a specific circuit with each operator of the language.
This circuit is similar to sequential boolean circuits except that wire values are
elements of $\xi$ boolean algebra. As a consequence, the equation system associated
with such a circuit handles $\xi$ valued variables. As already mentioned, solutions of equation
system allow to determine all signal status .

\begin{figure}[htbp]
\centerline{\epsfxsize=7cm \epsfbox{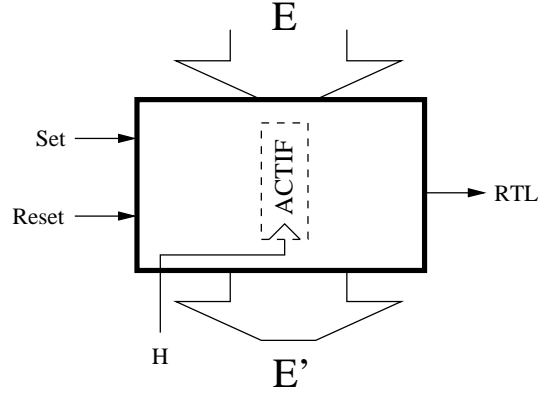}}
\caption{Circuit semantic for a \lstrl statement}
\label{circuit-semantic}
\end{figure}

To express the semantic of each statement in \lstrl, we generate a circuit whose
interface handles the following wires to propagate information and so to ensure synchronization 
between statements.
\begin{itemize}
\item \textit{SET} to propagate the control (input wire);
\item \textit{RESET} to propagate reinit (input wire);
\item \textit{RTL} ready to leave wire to indicate that the statement can terminate
in the reaction (or in a further one);
\end{itemize}

Wires used to synchronize sub programs are never
equal to $\bot$ or $\top$. They can be considered as boolean and the only values they can
bear are true or false. Thus, according to our translation from $\xi$ to
$\Bool \times \Bool$ :  ${\rm SET}_{def} = {\rm RESET}_{def}= {\rm RTL}_{def} = 1$.
In the following, we will denote $P_S$ the set of synchronization wires of $P$.
Moreover, for statements that do not terminate instantaneously, a register
is introduced (called \textit{ACTIF}). 
Similarly to control wires, ${\rm ACTIF}_{def} = 1$.
We will denote $P_R$ the set of registers of a program $P$.

In order to define the equational semantic, we introduce an operator: $\boxtimes$ that acts
on the element of $\xi$ whose boolean definition value is 1. Let $\xi_{\Bool} = \{x \in \xi | x_{def} = 1 \}$:
\[
\begin{array}{l l c l}
\boxtimes:& \xi \times \xi_{\Bool} &\longrightarrow &\xi_{\Bool}\\
          & (x,y) & \longrightarrow & (1, x_{def}.x_{val}.y_{val})
\end{array}
\]
This new operation will be useful to define the product between a real $\xi$ valued signal and
a synchronization wire or register. It is different from $\blacktriangleleft$ operation, since
this latter defines a ``mux'' operation and not a product.

In addition, we introduce  a ${\cal P}re$ operation on environment
in order to express the semantic of operators that do not react instantaneously. 
It allows to memorize all the status of current instances of events.
As already said, an environment is a set of events, but  circuit semantic handles wider 
environments than behavioral  semantic. In the latter, they contain only input and output
events, while in equational semantic they also contain event duplication and wires and registers.
Let $E$ be an environment, we denote $E \restr{I}$ the input events of $E$ and
$E\restr{O}$ the output ones.

\[
{\cal P}re(E) = \{S^\bot | S^x \in E ,\ S\not\in E\restr{I},
                                                \ S\not\in E\restr{O} \} \cup
                \{S_{pre}^x | S^x \in E,\ S\not\in E\restr{I},
                                                 \ S\not\in E\restr{O} \}
\]

The ${\cal P}re(E)$ operation consists in a duplication of events in the environment.
Each event $S^x$ is recorded in a new event $S_{pre}^x$ and the current value of signal
$S$ is set to $\bot$ in order to be refined in the current computation. But,
${\cal P}re(E)$ operation does not concern
interface signals because it is useless, only their value in the current instant is relevant.
Moreover, this operation updates the registers values:
we will denote ${\rm ACTIF}^+$ the value of the register ACTIF computed for the next reaction.

In \lstrl equational semantic, we consider $\xi$-circuits i.e circuits characterized
by a set of $\xi$-wires, a set of $\xi$-registers and an environment $E$ where
$\xi$-wires and $\xi$-registers have associated $\xi$ values.
The $\xi$-circuit schema is described in figure~\ref{circuit-semantic}. The $\xi$-circuit
{\footnote{
in what follow, when no ambiguity remains, we will omit the $\xi$ prefix when
speaking about $\xi$-circuit.
}}
associated with a statement  has an input environment E and generates and
output environment E'. The environment include input, output, local and
register status. 

We rely on the general theory of boolean constructiveness  previously detailed.
Let ${\cal C}$ be a $\xi$-circuit, we translate ${\cal C}$ into a boolean circuit.
More precisely, 
${\cal C} = (W, R, E)$ where 
$W$ is s set of $\xi$-wires  and $R$ a set of $\xi$-registers.
$E$ is composed of a set of equations of the form $ x = e$  in order to compute a
status for wires and registers.

Now, we translate ${\cal C}$ into the following boolean circuit ${\cal C}^B = (W^B,R^B,D^B)$
where $W^B$ is a set of boolean wires, $R^B$ a set of boolean registers and $D^B$ a set of 
boolean equations.

\[
\begin{array}{l}
W^B = \{ w_{def}, w_{val}\ |\ w \in W\} \\
R^B = \{ w_{def}, w_{val}\ |\ w \in R\} \\
D^B = \{ w_{def} = e_{def}, w_{val} = e_{val}\ | \ w = e \in E\} \\
\end{array}
\]

$e_{def}$ and $e_{val}$ are computed according to the algebraic rules detailed
section~\ref{bool-encoding}.

Now we define the constructive propagation law ($\rightsquigarrow$)
for $\xi$-circuits. Let ${\cal C}$ be a $\xi$-circuits with 
$I \subseteq E$ as input wire set and $R \subseteq E$ as register set,
the definition of the constructive propagation law for ${\cal C}$ is:
\[
E \vdash w \rightsquigarrow bb \Leftrightarrow I ^B,R^B \vdash w_{def} \hookrightarrow 
bb_{def\ } {\rm and}\ I ^B,R^B \vdash w_{val} \hookrightarrow bb_{val}.
\]

This definition is the core of the equational semantic. We rely on it to compile
\lstrl programs into boolean equations. Thus, we benefit from BDD representation and
optimizations to get an efficient compilation means. Moreover, we also rely on BDD
representation to implement a separate compilation mechanism.

Given $P$ a \lstrl statement.
Let ${\cal C}(P)$ be its  associated circuit 
{\footnote{ the equations defining its SET, RESET and RTL wires and the equations defining its registers
when it has some}}
and $E$ be  an input environment.
A reaction for the circuit semantic corresponds to the computation of an output environment
composition of $E$ and the synchronization equations of $P$. We denote $\bullet$ this composition
operation:
\begin{center}
$E' = E \bullet {\cal C}(P)$
if and only if
$E \cup {\cal C}(P) \vdash w \rightsquigarrow bb,\ {\rm and}\ E '(w) = bb, 
                                         \forall w \in E \cup {\cal C}(P)$.
\end{center}

Now, we define the circuit semantic for each statement of \lstrl. We will denote:
${\eqsem P E}$ the output environment of $P$ built from  $E$ input environment.

\subsection{Equational Semantic of LE Statements}

\subsubsection*{Nothing}

The circuit for \fixed{nothing} is described in figure~\ref{nothing} in appendix \ref{lecircuit}.
The corresponding equation system is the following:

{\small 
\[
{\eqsem {\rm nothing} {\rm E}} =  {\rm E} \bullet \{{\rm RTL} = {\rm SET} \}
\]
}

\subsubsection*{Halt}

The circuit for \fixed{halt} is described in figure~\ref{halt} in appendix \ref{lecircuit}. 
The statement is never ready to leave instantaneously.

{\small 
\[
{\eqsem {\rm halt} {\rm E}}   =  {\rm E}  \bullet \{ {\rm RTL} = 0 \}
\]
}

\subsubsection*{Emit}

The \fixed{emit} $S$ statement circuit is described in
figure~\ref{emit} in appendix \ref{lecircuit}. As soon as the
statement receive the control, it is ready to leave.   RTL and SET wires are equal  and
the emitted signal $S$ is present in the output environment. We don't
straightly put the value of $S$ to 1 in the environment, we perform a
\Plus operation with 1 in order to keep the possible value $\top$ and then transmit errors.
Moreover, the latter is
driven with the boolean value of RTL wire:

{\small
\[
{\eqsem {{\rm emit\ } S} {\rm E}} =
          ({\rm E}[S \leftarrow (1 \boxplus \xi(S))]) \blacktriangleleft {\rm RTL}_{val} 
\bullet \{ {\rm RTL}  = {\rm SET} \}
\] 
}

\subsubsection*{Pause}

The circuit for \fixed{pause} is described in figure~\ref{pause}  in appendix \ref{lecircuit}.
This statement does not
terminate instantaneously, as a consequence a register is created and a ${\cal P}re$ operation
is applied to the output environment:

{\small
\[
{\eqsem {\rm pause}{\rm E}}  = {\cal P}re({\rm E}) \bullet \left\{
\begin{array}{l c l}
{\rm RTL} & = & {\rm ACTIF}\\
{\rm ACTIF}^+ & = & ({\rm SET}\boxplus {\rm ACTIF}) \boxtimes\: \neg{\rm RESET}
\end{array}
\right\}
\] 
}

\subsubsection*{Wait}

The circuit for \fixed{wait} is described in figure~\ref{wait} in appendix \ref{lecircuit}.
The \fixed{wait} $S$ statement
is very similar to the \fixed{pause} one, except that the ready to leave wire is drive
by the presence of the awaited signal:

{\small 
\[
{\eqsem {{\rm wait\ } S} {\rm E}}  = {\cal P}re({\rm E}) \bullet \left\{
\begin{array}{l c l}
{\rm RTL} & = & {\rm ACTIF} \boxtimes S\\
{\rm ACTIF}^+ & = & ({\rm SET} \boxplus {\rm ACTIF}) \boxtimes \neg{\rm RESET} 
\end{array}
\right\}
\]
}

\subsubsection*{Present}

The circuit for ${\fixed Present}\ S \{{\rm P}_1\} \fixed{else} \{{\rm P}_2\}$ is described in 
figure~\ref{present}  in appendix \ref{lecircuit}.
Let E be an input environment, the SET control
wire is propagated to the then operand  ${\rm P}_1$ assuming signal S
is present while it is propagated to the else operand ${\rm P}_2$ assuming that S is absent.
The resulting environment E' is the \Plus law applied to the
respective outgoing environments  of ${\rm P}_1$ and ${\rm P}_2$.
Let E' be ${\eqsem {{\fixed Present}\ S \{{\rm P}_1\} \fixed{else} \{{\rm P}_2\}} {\rm E}}$,  
E' is defined as follows:

{\small 
\[
{\rm E'} =
\left[
\begin{array}{l}
  {\eqsem {P_1} E} \blacktriangleleft (S_{def}.S_{val})\boxplus \\
  {\eqsem {P_2} E} \blacktriangleleft (S_{def}.\overline{S_{val}})\boxplus\\
  {\rm E} \blacktriangleleft (\overline{S_{def}}.\overline{S_{val}})\boxplus\\
  {\rm E}^{\top} \blacktriangleleft(\overline{S_{def}}.S_{val})
\end{array} 
\right] \bullet \left\{
\begin{array}{l c l}
{\rm SET}_{{\rm P}_1} & =  &  {\rm SET} \blacktriangleleft (S_{def}.S_{val})\\
{\rm SET}_{{\rm P}_2} &=  &   {\rm SET} \blacktriangleleft (S_{def}.\overline{S_{val}})\\
{\rm RESET}_{{\rm P}_1} & = & {\rm RESET}\\
{\rm RESET}_{{\rm P}_2} & = & {\rm RESET}\\
{\rm RTL} & = &  {\rm RTL}_{{\rm P}_1}\boxplus {\rm RTL}_{{\rm P}_2}\boxplus
                 (1 \blacktriangleleft \overline{S_{def}}.S_{val})
\end{array}
\right\}
\]
}

\subsubsection*{Parallel}

Figure~\ref{parallel}  in appendix \ref{lecircuit} shows the circuit for ${\rm P}_1 \| {\rm P}_2$. 
The output environment contains the
upper bound of respective events in the output environments of ${\rm P}_1$ and ${\rm P}_2$.
The parallel is ready to leave when both ${\rm P}_1$ and ${\rm P}_2$ are:

{\small
\[
{\eqsem{P_1 \| P_2} E}  =  {\eqsem{P_1} E}\boxplus {\eqsem{P_2} E} \bullet \left\{
\begin{array}{lcl}
{\rm SET}_{{\rm P}_1} & =  & {\rm SET} \\
{\rm SET}_{{\rm P}_2} & =  & {\rm SET} \\
{\rm RESET}_{{\rm P}_1}& = & {\rm RESET} \\
{\rm RESET}_{{\rm P}_2} & = &{\rm  RESET}\\
{\rm ACTIF_1}^+ & = & ({\rm RTL}_{{\rm P}_1}\boxplus{\rm ACTIF}_1)\boxtimes \neg{\rm RESET}\\
{\rm ACTIF_2}^+ & = & {\rm RTL}_{{\rm P}_2}\boxplus{\rm ACTIF}_2)\boxtimes\neg{\rm RESET}\\
{\rm RTL} & = & ({\rm RTL}_{{\rm P}_1}\boxplus{\rm ACTIF}_1)  \boxtimes ({\rm RTL}_{{\rm P}_2}
                                                                  \boxplus {\rm ACTIF}_2) 
\end{array}
\right\} 
\]
}

\subsubsection*{Sequence}

Figure~\ref{sequence} in appendix \ref{lecircuit} shows the circuit
for ${\rm P}_1 \gg {\rm P}_2$.  The control is passed on from ${\rm
P}_1$ to ${\rm P}_2$: when ${\rm P}_1$ is ready to leave then ${\rm
P}_2$ get the control (equation 1) and ${\rm P}_1$ is reseted
(equation 2) :

{\small 
\[
{\eqsem{P_1 \gg P_2}E} = {\eqsem{P_1} E}\boxplus 
                         ({\eqsem{P_2} {\eqsem{P_1} E}} \blacktriangleleft {\rm RTL}_{{{\rm P}_1}_{val}})
\bullet \left\{
\begin{array}{l c l}
{\rm SET}_{{\rm P}_1} & =  & {\rm SET} \\
{\rm SET}_{{\rm P}_2} & =  & {\rm RTL}_{{\rm P}_1} (1) \\
{\rm RESET}_{{\rm P}_1} & = & {\rm RESET} \boxplus  {\rm RTL}_{{\rm P}_1}(2)\\
{\rm RESET}_{{\rm P}_2} & = & {\rm RESET} \\
{\rm RTL} & = & {\rm RTL}_{{\rm P}_2}
\end{array}
\right\}
\]
}

\subsubsection*{Abort}

The \fixed{abort} statement has for semantic the circuit described in figure~\ref{abort}
in appendix \ref{lecircuit}.
A register is introduced since the operator semantic is to not react instantaneously
to the presence of the aborting signal:

{\small
\[
{\eqsem {{\fixed abort}\ {\rm P}\ {\rm when}\ S} E}  =  {\eqsem {\rm P} E} \bullet \left \{
\begin{array}{l c l}
{\rm SET}_{\rm P} & =  & {\rm SET} \\
{\rm RESET}_{\rm P} & = & (\neg {\rm RESET}\boxdot S) \boxplus {\rm RESET}\\
{\rm RTL} & = & (\neg S \boxdot {\rm RTL}_{\rm P})\boxplus (S \boxtimes  
                                                             ({\rm SET}\boxplus {\rm ACTIF}))\\
{\rm ACTIF}^+ & = & ({\rm SET} \boxplus {\rm ACTIF}) \boxtimes \neg {\rm RESET}
\end{array}
\right\}
\]
}

\subsubsection*{Loop}

The statement \fixed{loop}\{P\} has for semantic the circuit described in figure~\ref{loop}
in appendix \ref{lecircuit}.
The \fixed{loop} statement does not terminate and similarly to its behavioral semantic,
its circuit semantic is equal to the one of P $\gg$ loop {P}:

{\small
\[
{\eqsem{{\fixed loop}\ {\rm P}} E} =  {\eqsem{\rm P} E} \bullet \left\{
\begin{array}{l c l}
{\rm SET}_{\rm P} & = & {\rm SET} \boxplus {\rm RTL}_{\rm P} \\
{\rm RESET}_{\rm P} & = & {\rm RESET} \\
{\rm RTL} & = & 0 \\
\end{array}
\right\}
\]
}

\subsubsection*{Local}

The \fixed{local} $S$ \{P\} statement restricts the scope of $S$ to
sub statement P. At the opposite to interface signals, such a signal
can be both tested and emitted. Thus, we consider that $S$ is a new
signal that does not belong to the input environment (it always
possible, up to a renaming operation). Let SET, RESET and RTL be the
respective input and output wires of the circuit, the equations of \fixed{local}
$S$ \{P\} are:

{\small
\[
{\eqsem {{\fixed local}\ {\rm S}\ \{ {\rm P} \}} E} = {\eqsem {\rm P} E}  \bullet \left\{
\begin{array}{l c l}
{\rm SET}_{\rm P} & =& {\rm SET}\\
{\rm RESET}_{\rm P} & = & {\rm RESET} \\
{\rm RTL} &=& {\rm RTL}_{\rm P} \\
S & = & \bot
\end{array}
\right\}
\]
}

\subsubsection* {Run\{P\}}

The circuit for \fixed{run}  statement is described
in figure~\ref{run}  in appendix \ref{lecircuit}.
Intuitively,  \fixed{run} \{P\} behaves similarly to P if P does not react instantaneously,
and to \fixed{pause} $\|$  P.
Thus, we get the following equation systems:

{\small
\[
{\eqsem {{\fixed run}\ {\rm P}} E}  = {\cal P}re({\rm E })\boxplus {\eqsem {\rm P} E} \bullet \left \{
\begin{array}{l c l}
{\rm SET}_{\rm P} & =& {\rm SET} \\
{\rm RESET}_{\rm P} & = & {\rm RESET} \\
{\rm ACTIF_1}^+ &=& ({\rm SET} \boxplus {\rm ACTIF}_1) \boxtimes \neg{\rm RESET}\\
{\rm ACTIF_2}^+ &=& ({\rm RTL}_{\rm P} \boxplus {\rm ACTIF}_2)\boxtimes \neg{\rm RESET}\\ 
{\rm RTL}  &=&  {\rm ACTIF}_1 \boxtimes ({\rm RTL}_{\rm P} \boxplus {\rm ACTIF}_2)\\
\end{array} \right \}
\]
}

\subsubsection* {Automata}

As already discussed, an automata is a finite set of \textit{macro
states}.  A macro state does not react
instantaneously, but takes at least an instant.  Figure~\ref{autom} in
appendix \ref{lecircuit} describes the circuit semantic for
${\cal A}({\cal M}, {\cal T}, {\cal C}ond, M_f, {\cal O}, \lambda)$.
The equational semantic of automata is the following set of equations:

{\small
\[
{\eqsem {\cal A} E} =  \displaystyle\sum_{M \in {\cal M}} {\eqsem {M}{{\cal E}_M}}\quad 
\bullet \bigcup_{M \in {\cal M}}\left\{
\begin{array} {l l l}
{\rm SET}_{\rm M} & = & 
    (\displaystyle\sum_{{\rm M}_i \rightarrow {\rm M}\in {\cal T}} {\rm RTL}_{{\rm M}_i} 
                          \blacktriangleleft c_{{\rm M}_i\rightarrow{\rm M}})
 \boxplus ({\rm SET} \blacktriangleleft \displaystyle\sum_{\rightarrow {\rm M} \in {\cal T}} 
                                                                       c_{\rightarrow {\rm M}}) \\
{\rm RESET}_{\rm M} & = & 
\displaystyle\sum_{{\rm M}\rightarrow {\rm M}_i \in {\cal T}}
      {\rm RTL}_{\rm M} \blacktriangleleft c_{{\rm M}\rightarrow{\rm M}_i}) \boxplus {\rm RESET} \\
{\rm RTL} & = & {\rm RTL}_{{\rm M}_{\rm f}}
\end{array}
\right\}
\]
}

where ${\cal E}_M$ is defined by:
{\small
\[
          E [s \leftarrow 1 \blacktriangleleft c_{\rightarrow M}\ | 
                                                    \ s \in \lambda(\rightarrow M)]
                   \blacktriangleleft \displaystyle\sum_{\rightarrow M \in {\cal T}}
                                                              c_{\rightarrow  M}) \: \boxplus 
          \displaystyle\sum_{ M_i \rightarrow M \in {\cal T}} 
              ( {\eqsem 
                  {M_i} 
                   {E[ s \leftarrow 1 \blacktriangleleft c_{M_i \rightarrow M}\ | \ s 
                                                            \in \lambda(M_i \rightarrow M)]}})
                   \blacktriangleleft c_{{M_i} \rightarrow M}
\]
}

To complete automata circuit semantic definition, we now detail the circuit for
macro states. Let M be a single macro state (which does not contain a \fixed{run} P instruction),
then its associated circuit is similar to the one of \fixed{pause}:

{\small
\[
{\eqsem {\rm M} {\rm E}} = {\cal P}re({\rm E}) \bullet \left\{
\begin{array}{lcl}
{\rm ACTIF}^+ & = & ({\rm SET}_{\rm M} \boxplus {\rm ACTIF}) \boxtimes \neg{\rm RESET}_{\rm M}\\ 
{\rm RTL}_{\rm M} & = & ACTIF\\
\end{array}
\right\}
\]
}

Otherwise, if the macro state M contains a {\fixed run} P instruction, its circuit is the combination of 
equations for single macro state and equations for {\fixed run} operator:

{\small
\[
{\eqsem {\rm M} {\rm E}} =   {\cal P}re({\eqsem {\rm P}{\rm E}}) \bullet \left\{
\begin{array}{lcl}
{\rm SET}_{\rm P} & =& {\rm ACTIF_1} \\
{\rm RESET}_{\rm P} & = & {\rm RESET}_{\rm M} \\
{\rm ACTIF_1}^+ &=& ({\rm SET}_{\rm M} \boxplus {\rm ACTIF}_1) \boxtimes \neg{\rm RESET}_{\rm M}\\
{\rm ACTIF_2}^+ &=& ({\rm RTL}_{\rm P} \boxplus {\rm ACTIF}_2)\boxtimes \neg{\rm RESET}_{\rm M}\\ 
{\rm RTL}_{\rm M} & = & {\rm ACTIF}_1 \boxtimes ({\rm RTL}_{\rm P} \boxplus {\rm ACTIF}_2)\\
\end{array}
\right\}
\]
}

Notice that a register is generated for each state, but in practice, we create only
${\rm log}_2 n$ registers if the automaton has $n$ states according to  the well-known
binary encoding of states.

\subsection{Equivalence between Behavioral and Circuit Semantic}

The circuit semantic allows us to compile \lstrl programs in a compositional way. 
Given a non basic statement $p\ {\rm Op}\ q$ (let Op be an operator of \lstrl),
then its associated circuit is deduced from
${\eqsem{p}{E}}$ and ${\eqsem {q} {E}}$ applying the semantic rules. On the other hand,
the behavioral semantic gives a meaning to each program and is logically correct,
and we prove now that these two semantic agree  on both the set of emitted signals and 
the termination  flag value for a \lstrl program $P$.
To prove this equivalence, we consider a global input environment $E$ containing
input events and output events set to $\bot$. Considering the circuits semantic,
the global environment (denoted $E_{\cal C}$) is $E \cup P_S \cup P_R$. 

To prove the equivalence between  behavioral and circuit semantic, first we introduce
a notation: let $P$ be a \lstrl statement, $SET(P)$, $RESET(P)$ and $RTL(P)$ will denote
respectively the SET, RESET and RTL wires of $P$. Second, we introduce the notion
of size for a statement.

\subsubsection*{Definition}

We define $\lceil P \rceil$, the size of P  as follows:

\begin{itemize}
\item $\lceil \fixed{nothing} \rceil$ = 1;
\item $\lceil\fixed{ halt} \rceil$ = 1;
\item $\lceil \fixed{emit} \rceil$ = 1;
\item $\lceil \fixed{pause} \rceil$ = 1;
\item $\lceil \fixed{wait} \rceil$ = 1;
\item $\lceil \fixed{present}\ \{P_1\}\ \fixed{else}\ \{P_2\} \rceil$ = 
$\lceil P_1\rceil$ + $\lceil P_2\rceil$ +1;
\item $\lceil P_1 \| P_2 \rceil$ =  $\lceil P_1\rceil$ + $\lceil P_2\rceil$ +1;
\item $\lceil P_1 \gg P_2 \rceil$ =  $\lceil P_1\rceil$ + $\lceil P_2\rceil$ +1;
\item $\lceil \fixed{abort}\ \{P\}\ \fixed{when}\ S \rceil$ = $\lceil P \rceil$ +1;
\item $\lceil \fixed{loop}\ \{P\} \rceil$ = $\lceil P \rceil$ +1;
\item $\lceil \fixed{local} S \ \{P\} \rceil$ = $\lceil P \rceil$ +1;
\item $\lceil \fixed{automata}({\cal M}, {\cal T}) \rceil$ = 
              $\sum \lceil M_i \rceil\ {\rm such\ that}\ M_i \in {\cal M}$ +1;
\end{itemize}

\newtheorem*{theorem}{Theorem}
\label{equiv-theorem}
\begin{theorem}

Let $P$ be a \lstrl statement and $E_{\cal C}$ an input environment,
For each reaction, the following property holds:

$\Gamma(P) \xrightarrow[E]{E ', {\rm TERM}(P)_{val}} \Gamma(P)'$,
where $E =  E_{\cal C} - \{w | w \in P_S\ {\rm or}\ w \in P_R \}$;
$TERM = RTL(P)_{val}$;
and ${\eqsem {P} {E_{\cal C}}} \restr{O} = E ' \restr{O}$
\end{theorem}

\subsubsection*{Proof}
 
We perform an inductive proof on the size of $P$. Notice that the
proof requires to distinguish the initial reaction from the others. In 
this reaction, $SET(P)=1$ and it is the only instant when this
equality holds. For statement reacting instantaneously, we consider only an
initial reaction  since considering following reactions is meaningless for them.

\subsubsection* {$\lceil P \rceil$ = 1},
We perform a proof by induction on the length of $P$. First, we prove the theorem for
basic statements whose length is 1.
According to the previous definition of $\lceil \rceil$, $P$ is either
\fixed{nothing}, \fixed{ halt}, \fixed{emit}, \fixed{pause} or
\fixed{wait}. 

\begin{enumerate}
\item $P = \fixed{nothing}$;

then  $\Gamma(P)$ = \fixed{nothing}. Following the equational semantic for \fixed{nothing}
statement:
\[ {\eqsem{P}{E_{\cal C}}} =  E_{\cal C} \bullet \{RTL(P) = SET(P)\} \]
Hence,
${\eqsem{P}{E _{\cal C}}}\restr{O}$ = 
${E_{\cal C}}\restr{O}$ = $E\restr{O}$ = $E'\restr{O}$.
Moreover, $RTL(P) = SET(P) = 1$ thus $RTL(P)_{val} = 1$;

\item $P$ = \fixed{halt};

then  $\Gamma(P)$ = \fixed{halt}. Similarly to \fixed{nothing},
${\eqsem {P}{E_{\cal C}}}\restr{O}$ = $E '\restr{O}$ and 
$RTL(P) = 0$ thus $RTL(P)_{val}$ = $0$;

\item $P$ = \fixed{emit} $S$;

then $\Gamma(P)$ = !$S$. 
As well in the behavioral rule for \fixed{!} as in the circuit equations for \fixed{emit},
we set the status of signal $S$ to 1 in the respective environments. From the definition,
$E_{\cal C} \restr{O} = E \restr{O}$ thus obviously, ${\eqsem{P}{E_{\cal C}}} \restr {O} =
E' \restr{O}$.
Moreover, $RTL(P) = SET(P) = 1$ thus, $RTL(P)_{val} = 1$.

\item $P$ = \fixed{wait} $S$;

According to the circuit semantic,
${\cal C}(P)$ has a register wire and we denote it $ACTIF(P)$.
The equations  for \fixed{wait} are:

{\small 
\[
{\eqsem{P}{E}}   = {\cal P}re({\rm E}) \bullet \left\{
\begin{array}{l c l}
RTL(P) & = & ACTIF(P) \boxtimes S\\
ACTIF(P)^+ & = & (SET(P) \boxplus ACTIF(P)) \boxtimes \neg RESET(P) 
\end{array}
\right\}
\]
}

The proof  of the theorem falls into two cases:
\begin{enumerate}
\item ACTIF(P)=0, we are in the initial reaction and then  $SET(P)=1$ , $RESET(P)=  0$.
it is obvious that $ACTIF(P)^+ = 1$. Then $ACTIF(P)$ becomes 1 in the 
environment according to the ${\cal P}re$ operation and all output wires keep their status
in $E '_{\cal C}$. 
When such a reaction occurs, 
in the behavioral semantic definition, the $wait$ rule is applied.
Following this rule  $E ' = E $. Thus,
${\eqsem {P}{E_{\cal C}}}\restr{O} = {E _{\cal C}}\restr{O}  = E\restr{O} = E '\restr{O}$, according to 
the ${\cal P}re$ operation definition which does not concern output signals.
From the equations above, we get $RTL(P) = 0$ whatever the status of $S$ is
and then $RTL(P)_{val} = 0$; this is in
compliance with the $wait$ rule.
Another situation where $ACTIF(P)=0$ is when $RESET(P)$ has been set to 1 
in the previous reaction.
This case occurs only if the wait statement is the first part of a $\gg$ operator
or the internal statement of an \fixed{abort} operator.
In both  cases, $RTL(P) = 0$ then $RTL(P)_{val} = 0 = TERM_{\Gamma(P)}$ and in both semantic
the outgoing environments remain unchanged and then the theorem still holds.

\item  ACTIF(P) = 1. we are not in the  initial reaction. 
Then, the corresponding rules applied in behavioral semantic are
either $iwait1$ or $iwait2$ depending of $S$ status in the environment.
Similarly to item 1, neither $iwait1$ and $iwait2$ rules nor ${\cal P}re$ operation 
change environment output signals, thus 
${\eqsem {P}{E_{\cal C}}}\restr{O} = E '\restr{O}$.

If $S^1 \in E_{\cal C}$ then $S^1 \in E $ since it is either an input signal
or a local one for the statement  and  then we apply rule $iwait1$, then 
$RTL(P) = 1$ and thus $RTL(P)_{val} = 1$. 
Otherwise, if $S^x \in E_{\cal C},\ x\not = 1$ and
$ACTIF(P)\:\boxtimes\: S = (1, S_{def}.S_{val}.ACTIF(P)_{val}) = (1,0)$ 
for $S = 0, \bot\ {\rm or}\ \top$.
Thus $RTL(P) = 0$  and  $RTL(P)_{val} = 0$. 
\end{enumerate}

\end{enumerate}

\subsubsection*{$\lceil P \rceil$ = n}
Now we study the inductive step, Assume that the theorem holds for statement whose length is
less than n. We study the case where the size of $P$ is n.
Then $P$ is either \fixed{present}, $\|$, $\gg$, \fixed{abort}, \fixed{loop}, \fixed{local} or
\fixed{automata} statement.

\begin{enumerate}
\item $P = \fixed{present}\:S\:\{P_1\}\ \fixed{else}\ \{P_2\}$;

Thus, according to the equational semantic, we know that: 

{\small
\[ 
\left \{
\begin{array}{l}
{\eqsem{P_1}{E_{\cal C}}} \blacktriangleleft (S_{def}.S_{val}) \:\boxplus \\
{\eqsem{P_2}{E_{\cal C}}} \blacktriangleleft (S_{def}.\overline{S_{val}}) \: \boxplus\\
E_{\cal C} \blacktriangleleft  (\overline{S_{def}}.\overline{S_{val}}) \:\boxplus\\ 
{E_{\cal C}}_{\top} \blacktriangleleft  (\overline{S_{def}}.S_{val})
\end{array}
\right \}
\bullet \{
RTL(P) =   RTL(P_1) \boxplus RTL(P_2) 
      \boxplus (1 \blacktriangleleft \overline{S_{def}}.S_{val}) \}
\subset {\eqsem{P}{E_{\cal C}}}
\]
}

On the other hand, $\Gamma(P) = S\ ?p_1 : p_2$ where $p_1 = \Gamma(P_1)$ and
$p_2 = \Gamma(P_2)$. The behavioral semantic
relies on the four rules defined in section~\ref{present-behav}:

By induction , we know that ${\eqsem{P_1}{E_{\cal C}}}\restr{O} = E '_1 \restr{O}$
and ${\eqsem{P_2}{E_{\cal C}}} \restr{O} = E '_2 \restr{O}$ where $E'_1$ (resp $E'_2$)
is the output environment of $p_1$ (resp $p_2$) computed from $E$ input environment,
and $RTL(P_1)_{val} = TERM_{p_1}$ and $RTL(P_2)_{val} = TERM_{p_2}$.
To prove the theorem for present operator, we  study
the different possible status of $S$ in the input environment (common to both semantic).
\begin{enumerate}
\item If $S$ is present, then $S_{def} = 1$ and $S_{val} = 1$. 
For the output signal valuation,
since  $S_{def}.S_{val} = 1$ , from the induction hypothesis
we deduce that ${\eqsem{P}{E_{\cal C}}}\restr{O} = E'\restr{O}$.
Concerning the $RTL$ wire and termination flag, if we consider present operator equations,
since  $S_{def} = 1$ and $S_{val} = 1$, we deduce that $SET(P_1) = 1$ and $SET(P_2) = 0$.
Thus $RTL(P_2) = 0$ too: either $P_2$ has no register and then its $RTL$ value depends straightly
of the $SET$ value, or $P_2$ has a register. In this case, its $RTL$ value depends of register 
value, but this latter cannot be 1 while the $SET$ value is 0.
Thus, $RTL(P)_{val} = RTL(P_1)_{val} = TERM_{P_1} = TERM_{\Gamma(P)}$ with respect to 
rule $present0$ in the behavioral semantic.
\item  If $S$ is absent, the prove is similar with $S_{def}=1$ and $S_{val} = 0$ and according
to the fact that rule $present1$ is applied from the behavioral semantic.
\item If $S$ status is $\bot$, then $S_{def}=0$ and $S_{val} = 0$.
In this case ${\eqsem{P}{E_{\cal C}}} = E_{\cal C}$ and $E' = E$, thus the result concerning 
outputs is obvious by induction.
Concerning $RTL$ wires and termination flag, since $S_{def} = 0$ thus both $SET(P_1)$ and
$SET(P_2)$ are 0 and then also $RTL(P_1)$ and $RTL(P_2)$ are. Thus $RTL(P) = 0$ and
$RTL(P)_{val} = 0 = TERM_{\Gamma(P)}$ according to rule $present3$ from behavioral semantic.
\item If $S$ has status $\top$, then an error occurs and in both semantic all signals in the 
environment are set to $\top$. In this case, $RTL(P) = 1$ and according to rule $present4$
of behavioral semantic, $RTL(P)_{val} = TERM_{\Gamma(P)} = 1$.
\end{enumerate}

\item $P = P_1 \| P_2$;

Thus, equations for $P$ are the following:
{\small
\[
 {\eqsem{P}{E_{\cal C}}}  = {\eqsem{P_1}{E_{\cal C}}}\boxplus {\eqsem{P_2}{E_{\cal C}}} \bullet \left\{
\begin{array}{lcl}
SET(P_1) & =  & SET(P) \\
SET(P_2) & =  & SET(P) \\
RESET(P_1)& = & RESET(P) \\
RESET(P_2) & = & RESET(P)\\
ACTIF_1(P)^+ & = & (RTL(P_1) \boxplus ACTIF_1(P)) \boxtimes \neg RESET(P)\\
ACTIF_2(P)^+ & = & RTL(P_2) \boxplus ACTIF_2(P) \boxtimes\neg RESET(P)\\
RTL(P) & = & (RTL(P_1)\boxplus ACTIF_1(P))\: \boxtimes \\
&&(RTL(P_2)\boxplus ACTIF_2(P)) 
\end{array}
\right\} 
\]
}

In \ple process algebra, $\Gamma(P) = p_1 \| p_2$, where $p_1 = \Gamma(P_1)$ and
$p_2 = \Gamma(P_2)$.  We recall the $parallel$ rule of behavioral semantic  for $\|$:
\[
\frac{p_1 \xrightarrow[E]{E '_1,\ TERM_{p_1}}p'_2\quad ,
      \quad p_2 \xrightarrow[E]{E '_2,\ TERM_{p_2}}p'_2}
     {p \xrightarrow[E]{E '_1 \:\boxplus\: E '_2 ,\: TERM_{p_1} . TERM_{p_2}} p'_1\|p'_2}
\]

By induction , we know that ${\eqsem{P_1}{E_{\cal C}}}\restr{O} = E '_1 \restr{O}$
and ${\eqsem{P_2}{E_{\cal C}}} \restr{O} = E '_2 \restr{O}$ and
$RTL(P_1)_{val} = TERM_{p_1}$ and $RTL(P_2)_{val} = TERM_{p_2}$.

Both equational and behavioral semantic perform the same \Plus operation on the environments
resulting of the computation of the respective semantic on the two operands.
Thus, the result concerning the outputs is straightly deduced from the induction hypothesis.

Concerning the $RTL$ wire,
$(RTL(P_1) \boxtimes RTL(P_2))_{val} = (RTL(P_1)_{val} . RTL(P_2)_{val}$ by
definition of $\boxtimes$ operation and according to the fact that
$RTL(P_1)_{def}= RTL(P_2)_{def} = 1$,and by induction
$RTL(P)_{val} = TERM_{p_1}.TERM_{p_2} = TERM_{\Gamma(P)}$.

\item $P = P_1 \gg P_2$;

 The equations for $\gg$ operator are the following:
{\small 
\[
{\eqsem{P}{E_{\cal C}}} = {\eqsem{P_1}{E_{\cal C}}} \boxplus 
                           ({\eqsem{P_2}{\eqsem{P_1}{E_{\cal C}}}} \blacktriangleleft RTL(P_1))
 \bullet \left\{
\begin{array}{l c l}
SET(P_1) & =  & SET(P) \\
SET(P_2) & =  & RTL(P_1) (1) \\
RESET(P_1) & = & RESET(P) \boxplus  RTL(P_1) (2)\\
RESET(P_2) & = & RESET(P) \\
RTL(P) & = & RTL(P_2)
\end{array}
\right\}
\]
}

In \ple process algebra, $\Gamma(P) = p_1 \gg p_2$ 
where $p_1 = \Gamma(P_1)$ and $p_2 = \Gamma(P_2)$.

The proof depends of the value of $RTL(P_1)$ in the equational semantic:
\begin{enumerate}
\item $RTL(P_1) = 0$;

By induction we know that $p_1 \xrightarrow[E]{E '_1, TERM_{p_1}} p'_1$ and
$TERM_{p_1} = RTL(P_1)_{val} = 0$. Then, in the behavioral semantic , rule $sequence1$ is
applied. Thus, $TERM_{\Gamma(P)} = 0$ and $E ' = E'_{p_1}$.
In the equational semantic, $SET(P_2) = RTL(P_1)$ thus $SET(P_2) = 0$ and so is
$RTL(P_2)$ (see the proof of \fixed{present} operator) and  $RTL(P)$ too.
$RTL(P_1)_{val} = 0$ and according to $\blacktriangleleft$ definition,
$({\eqsem{P_2}{\eqsem{P_1}{E_{\cal C}}}} \blacktriangleleft RTL(P_1))$ = $E_{\bot}$.
Thus, ${\eqsem{P}{E_{\cal C}}} = {\eqsem{P_1}{E_{\cal C}}} \bullet {\cal C}(P)$,
and ${\eqsem{P}{E_{\cal C}}} \restr{0} =  {\eqsem{P_1}{E_{\cal C}}} \restr{O}$.
On the other hand, in behavioral semantic, we have $E' \restr{O} = E'_{p_1}$.
Thus, from induction hypothesis, we deduce that:
${\eqsem{P}{E_{\cal C}}} \restr{O} = E' \restr{O}$.

\item $RTL(P_1) = 1$;

In this case, $TERM_{p_1} = RTL(P_1)_{val} = 1$ and rule $sequence2$ is applied in the
behavioral semantic.
By induction, we know that $TERM_{p_2} = RTL(P_2)_{val}$.
But, $RTL(P) = RTL(P_2)$ then $RTL(P)_{val} = TERM_p$. 
For  environments,
By induction, we also know that ${\eqsem{P_1}{E_{\cal C}}} \restr{O} = E'_{p_1} \restr{O}$.
In both semantic, the only way to change the value of an output signal in the environment is
with the help of the {\fixed emit} operator. Then, if the status of an output signal $o$ change
in ${\eqsem {P_1}{E_{\cal C}}}$ it is because $P_2$ involves an {\fixed emit} $o$ instruction.
Hence, relying on the induction hypothesis, we know that $o$ has the same status in 
${\eqsem {P_1}{E_{\cal C}}}$ and in $E'_{p_1}$. But, $o$ status cannot be changed in
two different ways in ${\eqsem {P}{E_{\cal C}}}$ and $E'$ since {\fixed emit} operator performs the
same operation on environments in both semantic.

\end{enumerate}

\item $P$ = \fixed{abort} $P_1$ \fixed{when} $S$;

Thus, the output environment is the solution of the following equations:
{\small
\[
{\eqsem{P}{E_{\cal C}}}  =  {\eqsem {P_1}{E_{\cal C}}} \bullet \left \{
\begin{array}{l c l}
SET(P_1) & =  & SET(P) \\
RESET(P_1) & = & (\neg (RESET(P)\boxdot S) \boxplus RESET(P)\\
RTL(P) & = & (\neg S \boxdot RTL(P_1))\boxplus (S \boxtimes (SET(P)\boxplus ACTIF(P)))\\
ACTIF(P)^+ & = & (SET(P) \boxplus ACTIF(P)) \boxtimes \neg RESET(P)
\end{array}
\right\}
\]
}

$\Gamma(P) = p_1 \uparrow s$ where $p_1 = \Gamma(P_1)$.

First, notice that in $abort1$, $abort2$ and $abort3$ of behavioral semantic, the output
environment $E'$ is $E'_{p_1}$. Similarly in the equational semantic 
$E'_{{\cal C}}$ is $E'_{{\cal C}_1}$ improved by the set of connexion wire equations for abort
statement. Then, applying the induction hypothesis, we
can deduce $E'\restr{O} = {\eqsem {P}{E_{\cal C}}} \restr{0}$.

Now, we prove that termination wire  coincides with termination flag in respective equational and
behavioral semantic.
We study first the case where we $S$ is present and then the case when it is not.
\begin{enumerate}
\item $S^1 \in E$;

Thus, $S^1 \in E_{\cal C}$ too.
In this case, $RTL(P) = SET(P) \boxplus ACTIF(P)$.
In the initial reaction $SET(P)=1$ and $ACTIF(P)=0$ and in further reaction
$SET(P) = 0$ and $ACTIF(P) = 1$. Then, in all reactions $RTL(P) = 1$.
On the other hand, it is rule $abort1$ that is applied in behavioral semantic and thus
$RTL_p = 1$. Hence, $RTL(P)_{val} = TERM_p$.
However, $ACTIF(P)$ can become 0. But, that means that in a previous reaction $RESET(P) = 1$
and $P$ is encompassed in a more general statement $P_g$ which is either another abort or a
sequence statement since there are the only operators that set the $RESET$ wire to 1.
If $P_g$ is an abort statement,
its abortion signal is 1 in the input environment and then we are in one of the 
previous case already studied. Otherwise, that means that $P$ is encompassed in the
first operand of $P_g$ whose $RTL$ is 1 and we can rely on the reasoning performed for sequence
operator to get the result we want. 

\item $S^1 \notin E$;
Thus, $S^1 \notin E_{\cal C}$ too.
If we expand the value of $S$ in the $RTL$ equation, we get $RTL(P)$ = $RTL(P_1)$.
In the behavioral semantic either rule $abort1$ or $abort2$ is applied
according to the value of $TERM_{p_1}$. But, whatever this value is, by induction we
get the result.

\end{enumerate}


\item $P$ = \fixed{loop} \{ $P_1$ \};

Thus,
{\small
\[
{\eqsem{P}{E_{\cal C}}} =  {\eqsem{P_1}{E_{\cal C}}} \bullet \left\{
\begin{array}{l c l}
SET(P_1) & = & SET(P) \boxplus RTL(P_1) \\
RESET(P_1) & = & RESET(P) \\
RTL(P) & = & 0 \\
\end{array}
\right\}
\]
} 
$\Gamma(P)$ = $\Gamma(P_1)*$  and rule $loop$ is applied in
the behavioral semantic to compute the reaction of $\Gamma(P)$.
According to this latter,  $p_1* \xrightarrow [E] {E '_1, 0} p'_1 \gg p_1*$ when
$p_1 \xrightarrow [E] {E '_1, TERM_{p_1}} p'_1$. By induction, we know that
${\eqsem{P_1}{E_{\cal C}}} \restr{O} = E '_{p_1} \restr{O}$ thus 
$ {\eqsem{P}{E_{\cal C}}} = E '\restr{O}$
and $RTL(P) = 0$ thus $RTL(P)_{val} = 0 = TERM_{\Gamma(P)}$.


\item $P$ = \fixed{local} $S$ \{$P_1$\};

According to the equational semantic, the following equations defined the local operator:
{\small
\[
{\eqsem{P}{E_{\cal C}}} = {\eqsem {P_1}{E_{\cal C}}}  \bullet \left\{
\begin{array}{l c l}
SET(P_1) & =& SET(P)\\
RESET(P_1) & = & RESET(P) \\
RTL(P) &=& RTL(P_1)\\
S & = & \bot
\end{array}
\right\}
\]
}
In \ple process algebra $\Gamma(P) = p_1 \backslash S$ where $\Gamma(P_1) = p_1$.
$local$ rule is applied  in the behavioral semantic:
$\Gamma(P) \xrightarrow[E]{E '_1 - \{S\}, TERM_{p_1}} p'_1 \backslash S$ when
$p_1 \xrightarrow [E \cup \{S\}] {E '_1, TERM_{p_1}} p'_1$.
Following the induction hypothesis, ${\eqsem{P_1}{E_{\cal C}}} \restr{O} = E '_{p_1} \restr{O}$ 
and $RTL(P_1)_{val} = TERM_{p_1}$.

Then ${\eqsem{P_1}{E_{\cal C}}}- \{S\} \restr{O} = E '_{p_1} \restr{O} - \{S\}$ and
${\eqsem{P}{E_{\cal C}}} \restr{O} = E ' \restr{O}$,
straightly from the induction hypothesis.


\item $P$ = \fixed{run} \{$P_1$\}

The \fixed{run} operator is not a primitive one, and we defined it as:
${\rm wait}\ \texttt{tick} \gg P_1$. Thus, the property holds for \fixed{run} operator
since it holds for both wait and $\|$ operators.


\item $P = {\cal A}({\cal M}, {\cal T}, {\cal C}ond, M_f, {\cal O}, \lambda)$.

Automata are both terms in \ple process algebra and programs in \lstrl language.
The equations for automata are the following:

{\small
\[
{\eqsem {P} {E_{\cal C}}} =  \displaystyle\sum_{M \in {\cal M}} {\eqsem {M}{{\cal E}_M}}\quad 
\bullet \bigcup_{M \in {\cal M}}\left\{
\begin{array} {l l l}
SET(M) & = & 
    (\displaystyle\sum_{M_i \rightarrow  M\in {\cal T}} RTL(M_i) 
                          \blacktriangleleft c_{M_i\rightarrow M})\quad  \boxplus\\
&& (SET(P) \blacktriangleleft \displaystyle\sum_{\rightarrow M \in {\cal T}} 
                                                                         c_{\rightarrow M}) \\
RESET(M) & = & 
\displaystyle\sum_{M \rightarrow M_i \in {\cal T}}
        RTL(M) \blacktriangleleft c_{M \rightarrow  M_i} \boxplus \\
&&  RESET(P) \\
RTL(P) & = & RTL(M_f)
\end{array}
\right\}
\]
}

where ${\cal E}_M$ is defined by:
{\small
\[
\begin{array} {l}
          E [s \leftarrow 1 \blacktriangleleft c_{\rightarrow M}\ | 
                                                    \ s \in \lambda(\rightarrow M)]
                   \blacktriangleleft \displaystyle\sum_{\rightarrow M \in {\cal T}}
                                                              c_{\rightarrow  M}) \: \boxplus \\
          \displaystyle\sum_{ M_i \rightarrow M \in {\cal T}} 
              ( {\eqsem 
                  {M_i} 
                   {E[ s \leftarrow 1 \blacktriangleleft c_{M_i \rightarrow M}\ | \ s 
                                                            \in \lambda(M_i \rightarrow M)]}})
                   \blacktriangleleft c_{{M_i} \rightarrow M}
\end{array}
\]
}

First of all, let us consider macro states. These latter are either
single macro states equivalent to a \fixed{pause} statement, or they
contains a \fixed{run} P instruction and then are equivalent to a
\fixed{pause} $\gg$ P instruction. In both cases, we have already
prove that the theorem holds.  Now, to prove the theorem for automata,
we perform an inductive reasoning on the sequence of reactions.

In the first reaction $SET(P) = 1$. All the $RTL(M)$ are 0, since macro states have at least a one
instant duration, thus $RTL(P)_{val} = 0$. On the other side, in behavioral semantic, rule
$automaton0$ is applied and $TERM_p = 0$ too.

For environments, for each macro states $M$, in the first reaction 
\[
{\cal E}_M = E_{\cal C} [s \leftarrow 1\ {\rm when}\ c_{\rightarrow M} =1\
                           {\rm and}\ s \in \lambda(\rightarrow M) \
                           {\rm and}\ {\rightarrow M \in {\cal T}}]
\]
When looking at equations related to $M$, we see that no output signal
status can be modified in the first reaction: either it is a single
macro state and then no output signal is modified whatever the
reaction is, or it contains a \fixed{run} $P_0$ statement but
$SET(P_0)$ cannot be true in the initial reaction and so no output
signal status can't be modified (the only operator that modified
output status is the \fixed{emit} one, and if the $SET$ wire of an
\fixed{emit} statement is not 1, the status of the emitted signal
remains unchanged). Thus, as well the behavioral semantic in rule
$automata0$ as the equational semantic set to 1 in their respective
environment output signal emitted in the initial transitions that
teach $M$. Hence, $E_{\cal C} \restr{O} = E \restr {O}$.

Now, we consider that the result is proved for the previous $n$
reactions, and we prove the result for the $n+1$ reaction.  In this
reaction, for each macro state $M \neq M_f$, if there is no transition
$M_i \rightarrow M$ such that $c_{M_i \rightarrow M}$ =1, then ${\cal
E}_M = E_{{\cal C}_n}$ where $E_{{\cal C}_n}$ in the environment
obtained after the previous $n$ reactions. Thus, ${\eqsem {M}{{\cal
E}_M}}$ is $E_{{\cal C}_n} \bullet {\cal C}(M)$ where ${\cal C}(M)$ is
the set of equations related to macro state $M$. In the behavioral
semantic, it is rule $automaton1$ which is applied and thus relying
both on induction hypothesis ensuring that $E_{{\cal C}_n} \restr{O} =
E_n \restr{O}$ and on the fact that the theorem is true for macro
state, we deduce the result.  Concerning the $TERM$ flag, as $M$ is not
final, $SET(M_f) = 0$ and so is $RTL(M_f)$. Hence, $RTL(P) = 0$, thus
$RTL(P)_{val} = 0$ too and $RTL_{\Gamma(P)}$ too (cf rule
$automaton1$).  On the other hand, if there is a transition $M_i
\rightarrow M$ such that $c_{M_i \rightarrow M} = 1$, ${\cal E}_M =
E_{{\cal C}_n} \boxplus {\eqsem {M_i}{E_{{\cal C}_n} [s \leftarrow 1\
| s \in \lambda(M_i \rightarrow M)]}}$.  If there is a transition $M_k
\rightarrow M$ such that $c_{M_k \rightarrow M}$ =1, thus similarly to
the case where $n$ = 1, we have
\[
{\cal E}_M =  E_{{\cal C}_n} [s \leftarrow 1\ {when}\  c_{M_k\rightarrow M} =1\ {\rm and}\ 
                       s \in  \lambda(M_k\rightarrow M) \ {\rm and}\ {\rightarrow M \in {\cal T}}]
\]
Since $E_{{\cal C}_n}$ is the resulting environment of the previous
instant, we know that $E_{{\cal C}_n} \restr{O}$ = $E_n \restr{O}$ On
the other hand, it is rule $automaton2$ that is applied in the
behavioral semantic and the output environment is modified in the
same way for both semantic. Similarly to the first instant, equations
for $M$ cannot modified the environment the first instant where
$SET_M$ is 1. Then, $E_{{\cal C}_{n+1}} \restr{O} = E_{n+1}
\restr{O}$.  In this case $RTL(M_f)$ is still 0 , thus $RTL(P) = 0$
and so is $RTL(P)_{val}$. Hence, according to rule $automaton2$,
result for $TERM$ flag holds.

Now, we will consider that there is a transition $M_k \rightarrow M_f$
such that $c_{M_k \rightarrow M_f} = 1$ {\footnote{the demonstration
is the same when the transition is initial (i.e $c{\rightarrow M_f}=1$)}}.  In this case, a
similar reasoning to the case where $M$ is not a final macrostate
concerning output environments holds. For termination flag, in equational
semantic, $RTL(P) =1$ when $RTL(M_f) = 1$. A similar situation holds
for behavioral semantic where it is rule $automata4$ which is
applied. Then, the result is deduced from the general induction
hypothesis since the size of macro states is less than the size of
automata from the definition.

\end{enumerate}

\hfill$\clubsuit$

\section {LE Modular Compilation}
\label{lecompilation}
\subsection{Introduction}
In the previous section, we have shown that every construct of the
language has a semantic expressed as a set of  $\xi$ equations. 
The first compilation step is the generation of a $\xi$ equation system
for each \lstrl program,
According to the semantic laws described in section \ref{leequational}.
Then, we translate each $\xi$ circuit into a boolean
circuit relying on the bijective map from $\xi$-algebra to $\Bool \times \Bool$
defined in section \ref{lebehavioral}.
This encoding allows us to translate $\xi$ equation system into a boolean
equation system (each equation being encoded by two boolean equations).
Thus, we can rely on a  constructive propagation law to implement equation system evaluation and
then, generate code, simulate or link with external code. But this approach
requires to find an evaluation order, valid for all synchronous
instants. Usually, in the most popular synchronous languages existing, this
order is static. This static ordering forbids any separate compilation mechanism
as it is illustrated in the following example. Let us consider the two
modules \texttt{first} and \texttt{second} compiled in a separate
way. Depending on the order chosen for sorting independant variables of each modules,
their parallel combination may lead to a causality problem (i.e there is a dependency cycle in the
resulting equation system).

\begin{figure}[htbp]
\centering
\includegraphics[width=11cm] {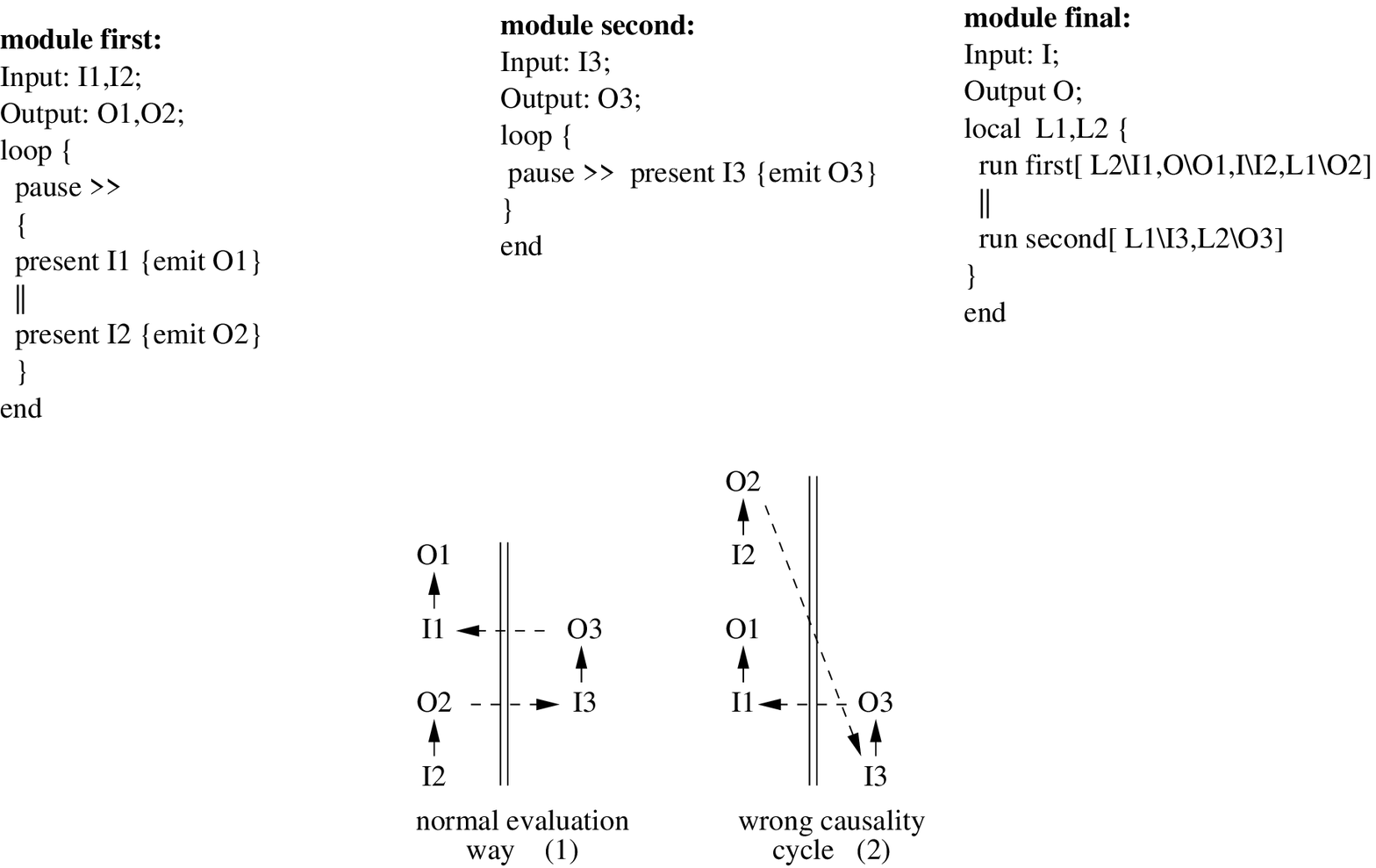}
\caption{Causality cycle generation: O1, O2 and O3 signals are independent. But, when choosing
a total order, we can introduce a causality cycle. If ordering (1) is chosen, in module final,
taking into account the renaming, we obtain the system: \{ L1 = I, L2 = L1, O = L2 \} which is
well sorted. At the opposite, if we choose ordering (2), in module final we get: 
\{ L2 = L1, O = L2,L1 = I \} which has a causality cycle.}
\label{fig:total-order}
\end{figure}

Figure~\ref{fig:total-order} describes a \lstrl module calling two sub modules.
Two compilation scenarios are shown on the right part of the figure. The
first one leads to a sorted equation system while the second
introduces a fake causality cycle that prevents any code
generation. 
Independent signals must stay not related: we aim at
building an incremental partial order.
Hence, while ordering the equation system, we keep enough
information on signal causality to preserve the independence of
signals. At this aim, we define two variables for each equation, namely
(\textit{Early Date}, \textit{Late Date}) to record the level when the
equation \textit{can} (resp. \textit{must}) be evaluated.  Each level
is composed of a set of independent equations. Level 0 characterizes
the equations evaluated first because they only depend of free
variables, while level n+1  characterizes the equation needed the evaluation of
variables from  lower levels (from n to 0) to be evaluated.
Equations of same level are independent and so can
be evaluated whatever the chosen order is.  This methodology is derived from the
\texttt{PERT} method. This latter
is well known for decades in the industrial production. Historically,
this method has been invented for the spatial conquest, back to the
60th when the NASA was facing the problem of synchronizing 30,000
independent, thus "concurrent", dealers to built the Saturne V rocket.
 
\subsection{Sort algorithm: a PERT family method}

Usually, the \texttt{PERT} method is applied in a task management context and each
task has a duration. In our usage, taking account duration of task makes no sense
 and the algorithm we
rely on to implement the \texttt{PERT} method is simplified. It is
divided into two phasis. The first step constructs a
forest where each tree represents variable dependencies.
Thus an initial partial order is built.  The second step is the 
recursive propagation of early and late dates. 
If during the propagation, a cycle 
is found there is a causality cycle in the program.
Of course the propagation ends since the number of
variables is finite. At worst, if the algorithm is successful (no causality cycle is found),
we can find a total order 
with a single variable per level (n variables and n levels). 
 
\subsubsection{Sorting algorithm Description}

More precisely, the first step builds two dependency sets
(\textit{upstream, downstream}) for each variable with respect to 
the equation which defines it. This first algorithm is detailed in appendix
\ref{pertalgorithm_1}.  The \textit{upstream} set of a variable $X$ is
the set of variables needed by  $X$ to be computed while the
\textit{downstream} set is the variables that need the value of $X$ to
be evaluated.  In practice, boolean equation systems are implemented
using binary decision diagrams (BDDs). Consequently the computation of
the downstream table is given for free by the BDD library.

\begin{figure}[htbp]
  \centering
      \subfigure[Equation system] {
      \includegraphics[width=2cm,ext=.eps]{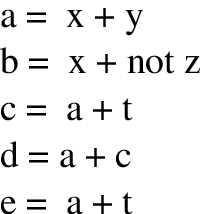}
      \label{fig::equation}
      }
      \subfigure[Dependences forest]{
      \includegraphics[width=5cm,ext=.eps]{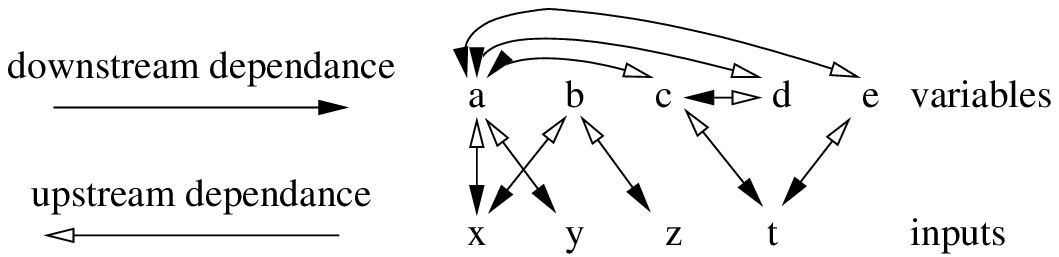}
      \label{fig::dependance-forest}
      }
      \subfigure[Date  propagation]{
      \includegraphics[width=5cm,ext=.eps]{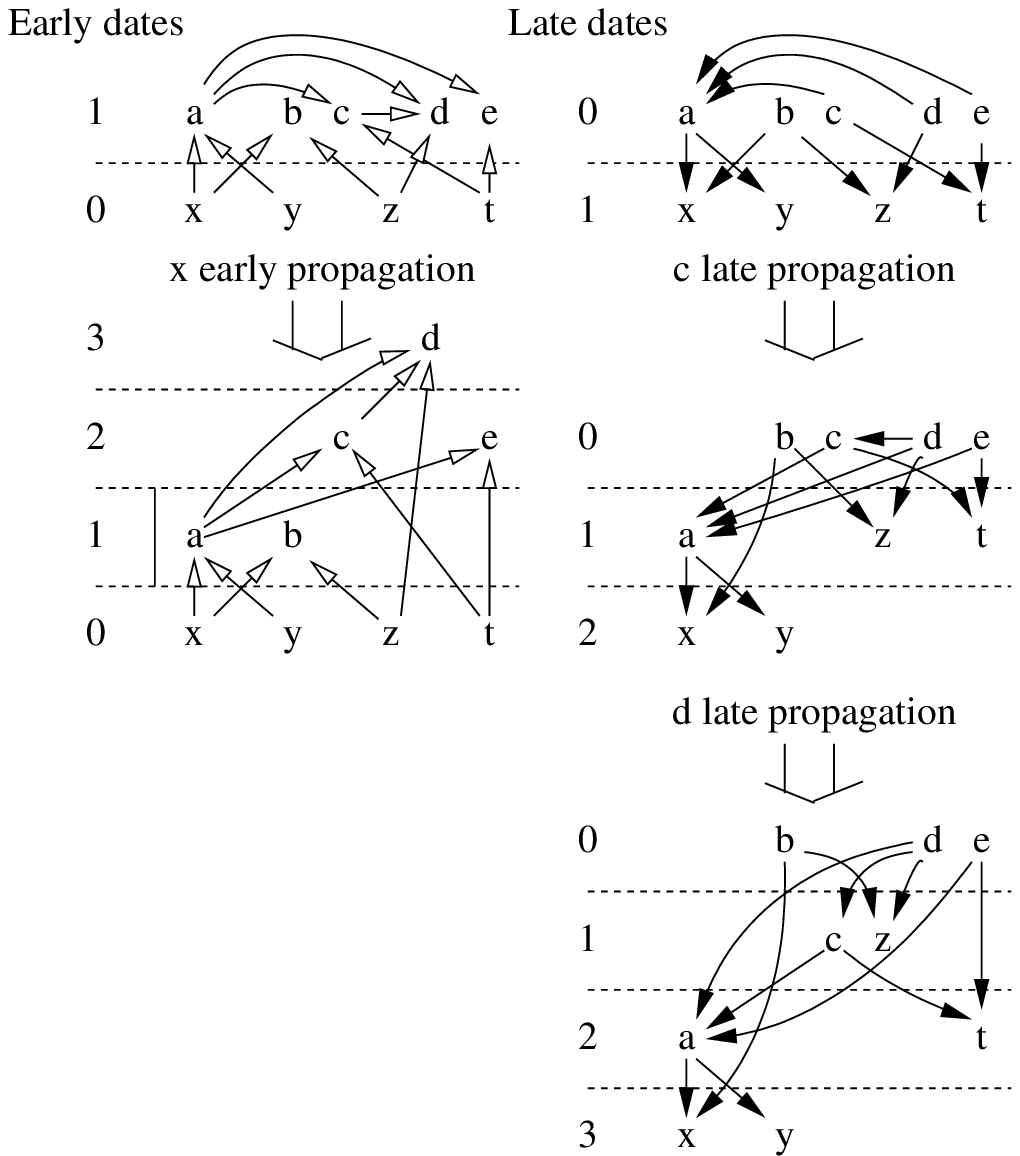}
      \label{fig::wood}
      }
     \caption{The dependence forest and propagation law application
              for a specific equation system.The different pert levels are specified 
              on the left hand side of figure~\ref{fig::wood}.}
\end{figure}  

We illustrate, the sorting algorithm we built on  an example.
Let us consider the set of equations expressed in figure~\ref{fig::equation}.
After the first step, we obtain the dependencies forest described in
figure~\ref{fig::dependance-forest},
Then, we perform early and late dates propagation.
Initially, all variables are considered independent and  their 
dates (\textit{early , late}) are  set  to ($0$,$0$). 
The second step  recursively propagates the \textit{Early}
Dates from the input and the register variables to the output variables and
propagates
the \textit{Late} Dates from the output variables to the input and the
register ones  according to a  \textit{n log n} propagation algorithm.
The algorithm that implements this second phasis is detailed in appendix \ref{pertalgorithm_2}.
Following  the example presented in figure~\ref{fig::equation}, the
algorithm results in the dependencies described in figure~\ref{fig::wood}.

\subsubsection{Linking two Partial Orders}

The approach allows an efficient merge of two already sorted equation
systems, useful to perform separate compilation.  To link the forest
computed for module~1 with the forest computed for module~2, we don't
need to launch again the sorting algorithm from its initial step.  In
fact, it is sufficient to only adjust the $early(late)\ dates$ of the
common variables to both equation systems and their
dependencies. Notice that the linking operation applies $\xi$-algebra
plus operator to merge common equations (i.e equations which compute
the same variable). Then, we need to adjust evaluation dates: every
output variable of module~1 propagates new $late\ date$ for every
downstream variables.  Conversely, every input variable of module~2
propagates new $early\ date$ for every  upstream variables.

\subsection{Practical Issues}

We have mainly detailed the theoretical aspect of our approach, and in this section we will discuss
the practical issues we have implemented.

\subsubsection{Effective compilation}

 Relying on the equational semantic, we compile a \lstrl program into
a $\xi$-algebra equation system. We call the compilation tool that
achieves such a task {\sc clem} 
(Compilation of LE Module).  In order to perform separate compilation
of \lstrl programs, we define an internal compilation format called
{\sc lec} (\lstrl Compiled code).  This format is highly inspired from
the Berkeley Logic Interchange Format ({\sc blif}~\footnote{\tt
http://embedded.eecs.berkeley.edu/Research/vis}). This latter is a
very compact format to represent netlists and we just add to it
syntactic means to record the \textit{early date} and \textit{late
date} of each equation. Practically, {\sc clem} compiler, among other
output codes, generates {\sc lec} format in order to reuse already
compiled code in an efficient way, thanks to the {\sc pert} algorithm
we implement.

\subsubsection{Effective Finalization}

Our approach to compile \lstrl programs into a sorted $\xi$ equation
system in an efficient way requires to be completed by what we call a
\textit{finalization} phasis to be effective.
To generate code
for simulation, verification or  evaluation, we must start from a \textit{valid}
boolean equation systems, i.e we consider only equation systems
where no event has value $\top$, since that means there is an error an
we propagate this value to each element of the environment in the
semantic previously described. Validity also means  well sorted equation
systems, to avoid to deal with programs having causality cycle. 
But in our approach we never set input event status to $absent$. 
Hence, we introduce a \textit{finalization} operation which  replaces all $\bot$ input events by
$absent$ events and propagates this information in all equations related to local variables and
outputs. Notice that the finalization operation is harmless. The
sorting algorithm relies on propagation of signal status, and the substitution
of $\bot$ by $absent$ cannot change the resulting sorted environment.

Let us illustrate the finalization mechanism on an example.
In the following code $O1$ and $O2$ depends on the I status:

\begin{verbatim}
loop {
  present I {emit O1} else {emit O2} 
  >> pause
}
\end{verbatim}

Before finalization, we get the following equation system:

\begin{tabular}{l}
$O1_{def}=I_{def}$ \\
$O1_{val}= I_{val}.I_{def}$ \\
$O2_{def}=I_{def} $  \\
$O2_{val}= \neg I_{val}.I_{def}$
\end{tabular}

We can see that
$O1_{def}$ and $O2_{def}$ are not constant because $I$ is not
necessarily defined for each instant (i.e $I_{def}$ can be $0$ if $I$ is $\bot$).
After finalization $I_{def}$ is set to $1$ and
$I_{val}$ remains free. According to the mapping from 
$\xi$ algebra to $\Bool \times \Bool$, an event $X$  such that $X_{def} = 0$ is either
$\top$ or  $\bot$. Since, we discard equation systems where an event has value $\top$,
To switch from $\bot$ value to $absent$ value, it is sufficient to set the $def$ part of a 
variable to 1.
Now for each logical instant the status ({\em
present}, {\em absent}) of I is known. The $O1$ and $O2$ equations
become:

\begin{tabular}{l}
$O1_{def}=1$ \\
$O1_{val}= I_{val}$ \\
$O2_{def}=1$  \\
$O2_{val}= \neg I_{val}$ 
\end{tabular}

We bring together compilation and finalization processus in a tool named {\sc clef}(Compilation of
LE programs and Finalization).

\subsubsection{Compilation scheme}
\label{compilation-scheme}

\begin{figure}[htbp]
\centering
\includegraphics[width=9cm] {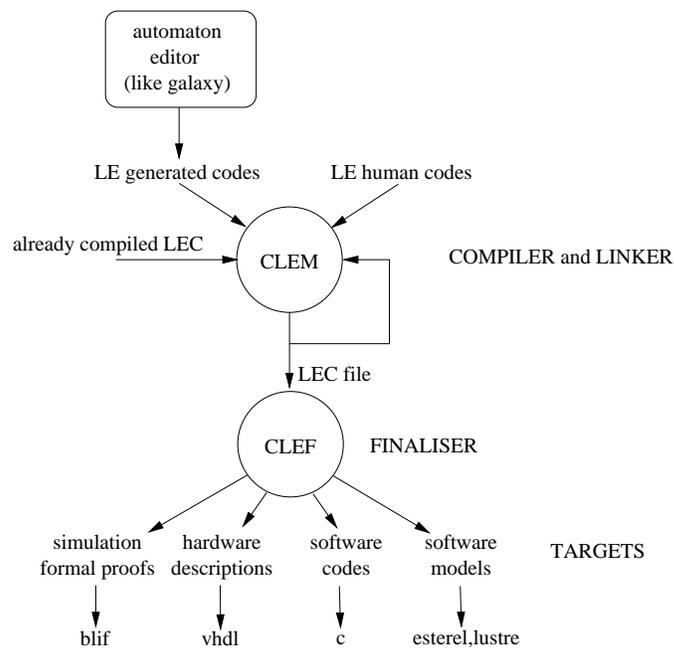}
\caption{Compilation Scheme}
\label{fig:compilation}
\end{figure}

Now, we detail the toolkit we have to specify, compile , simulate and execute
\lstrl programs.
A \lstrl file can be directly written. In the case of automaton,
it can be generated by automaton editor like \texttt{galaxy} too. Each
{\sc le} module is compiled in a {\sc lec} file and includes one
instance of the {\sc run} module references. These references can
be already compiled in the past by a first call of the \texttt{clem}
compiler. When the compiled process will done, the finalization will
simplify the final equations and generate a file in the target use:
simulation, safety proofs, hardware description or software code. That
is summed up in the figure~\ref{fig:compilation}.

\subsection {Benchmark}
\label{benchmark}
To complement  the experimentation of the example, we have done some tests about
the \textsc{clem} compiler. So we are interested in the evolution of  
the generated code enlarging with respect to the number of parallel processes increasing. A good 
indicator is the number of generated registers. Indeed, with $n$ registers, we can
implement $2^n$ states in an automaton.

The chosen process is very simpler, not to disturb the result:
\begin{verbatim}
module WIO:
Input: I; Output: O;
wait I >> emit O
end
\end{verbatim}    

which waits the I signal and emits the signal O one time as soon as I occurs. 
Here is the obtained table by the figure~\ref{fig:bench}:

\begin{figure}[htb]
\centering
\includegraphics[width=5cm]{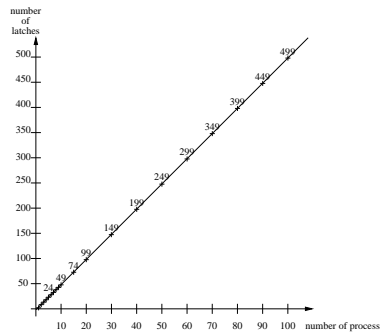}
\caption{Evolution of the Registers number}
\label{fig:bench}
\end{figure}

The relation between numbers of processes and number of registers seems to be linear, 
that is an excellent thing! The linear observed factor
of $5$ is only characterized by the equational semantic of 
{\tt parallel} and {\tt run} statements. In a next equationnal semantic, 
this number should be reduced.

\section {Example}
\label{leexample}

We illustrate \lstrl usage  on an industrial example concerning the design of a
mecatronics process control: a pneumatic prehensor. We first describe how the
system works. Then we present the system implementation with \lstrl language.
Finally, simulation  and verification are performed.

\subsection{Mecatronics System Description}

\begin{figure}[htbp]
\centering
\includegraphics[width=8cm] {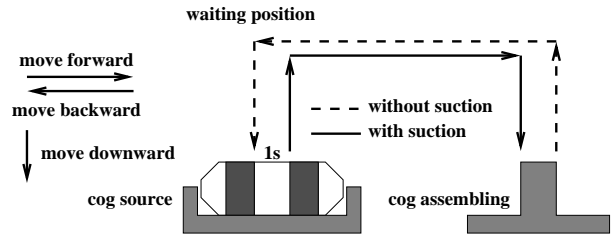}
\caption{A pneumatic prehensor}
\label{fig:cinema}
\end{figure}

A pneumatic prehensor takes and assembles  cogs and axes.
The physical system mainly consists of two 
double acting pneumatic cylinders and a suction pad. 
This example has been taken as a benchmark by an automation specialist 
group\footnote{{\tt http://www.lurpa.ens-cachan.fr/cosed}},
to experiment
new methods of design and analysis of {\em discrete event systems}. 
The (U cycle) kinematics of the 
system is described in Fig.\ref{fig:cinema}. 
Note that the horizontal motion must always be done in the
high position.

The horizontal motion pneumatic cylinder is driven by a 
bistable directional control valve (bistable {\sc dcv}). The associated
commands are \texttt{MoveFor} (short for move forward) and 
\texttt{MoveBack} (short for move backward). 
The vertical cylinder is driven by a monostable directional control valve 5/2 
whose active action is \texttt{MoveDown} (move downward).
In the absence of activation, the 
cylinder comes back to its origin position (high position). 
The suction pad (\texttt{SuckUp} command) is activated by a 
monostable {\sc dcv} (the suction is done by a Venturi effect).

\begin{figure}[htbp]
\centering
\includegraphics[width=6cm ]{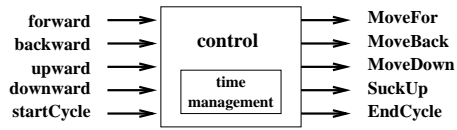}
\caption{Input/output signals}
\label{fig:es}
\end{figure}

\subsection {Mecatronics System LE Implementation}

In what follow we consider the control part of the system.
Fig.\ref{fig:es} gathers incoming information (from the limit switches
associated with the cylinders) and outgoing commands (to the
pre-actuators).  To implement this application in \lstrl language, we
adopt a top down specification technique. At the highest hierarchical
level , the controller is the parallel composition of an
initialization part followed by the normal cycle running and a
temporisation module.  This last is raised by a signal
\texttt{start\_tempo} and emits a signal \texttt{end\_tempo} when
the temporisation is over. Of course, these two signals are not in
overall interface of the controller, they are only use to establish
the communication between the two parallel sides.
The following \lstrl program implements the high level part of the
controller:

{\small
\begin{verbatim}
module Control:

Input:forward, backward, upward,  downward, 
      StartCycle;
Output:MoveFor, MoveBack, MoveDown, SuckUp, 
       EndCycle ;

Run: "./TEST/control/" : Temporisation;
     "./TEST/control/" : NormalCycle;

local start_tempo, end_tempo {
   { wait upward >> emit MoveFor 
     >> wait backward >> run NormalCycle
   }
 ||
   { run Temporisation}
}
end
\end{verbatim}
}

The second level of the specification describes temporisation and
normal cycle phasis.  Both Temporisation and NormalCycle modules are
defined in external files.  Temporisation module performs a
delaying operation (waiting for five successive reactions and then emitting
a signal \texttt{end\_tempo}. 
The overall \lstrl code is detailed in appendix \ref{lecontrolcode}.
In this section, we only discuss the
NormalCycle module implementation.  NormalCycle implementation is a
loop whose body specifies a single cycle.  According to the
specification, a single cycle is composed of commands to move the
pneumatic cylinders with respect to their positions and a call to a
third level of implementation (Transport) to specify the suction pad
activity.

{\small
\begin{verbatim}
module Transport :

Input: end_tempo, upward, forward, downward;
Output: MoveFor, MoveDown, SuckUp;

local exitTransport {

    { emit MoveDown >> wait end_tempo 
      >> wait upward >> emit MoveFor 
      >> wait forward >> emit MoveDown 
      >> wait downward >> emit exitTransport
    }
 ||

    abort
      { loop { pause >> emit SuckUp }}
    when exitTransport
}

end 

module NormalCycle :

Input:  StartCycle, downward, upward, backward, 
        forward, end_tempo;
Output: start_tempo, MoveDown, MoveBack, 
        MoveFor, SuckUp, EndCycle;

{ present StartCycle { nothing} else wait StartCycle}

>>

{
  loop { emit MoveDown 
         >> wait downward >> emit start_tempo  
         >> run Transport 
         >> wait upward >> emit MoveBack 
         >> wait backward >> emit EndCycle }
} 
end
\end{verbatim}
}

To compile the overall programs, we performed a separate compilation:
first, Temporisation and NormalCycle modules have been compiled and respectively saved in
\texttt{lec} format file. Second, the main Control module has been compiled according to
our compilation scheme (see figure~\ref{fig:compilation}).

\begin{figure}[htbp]
\centering
\includegraphics[width=10cm]{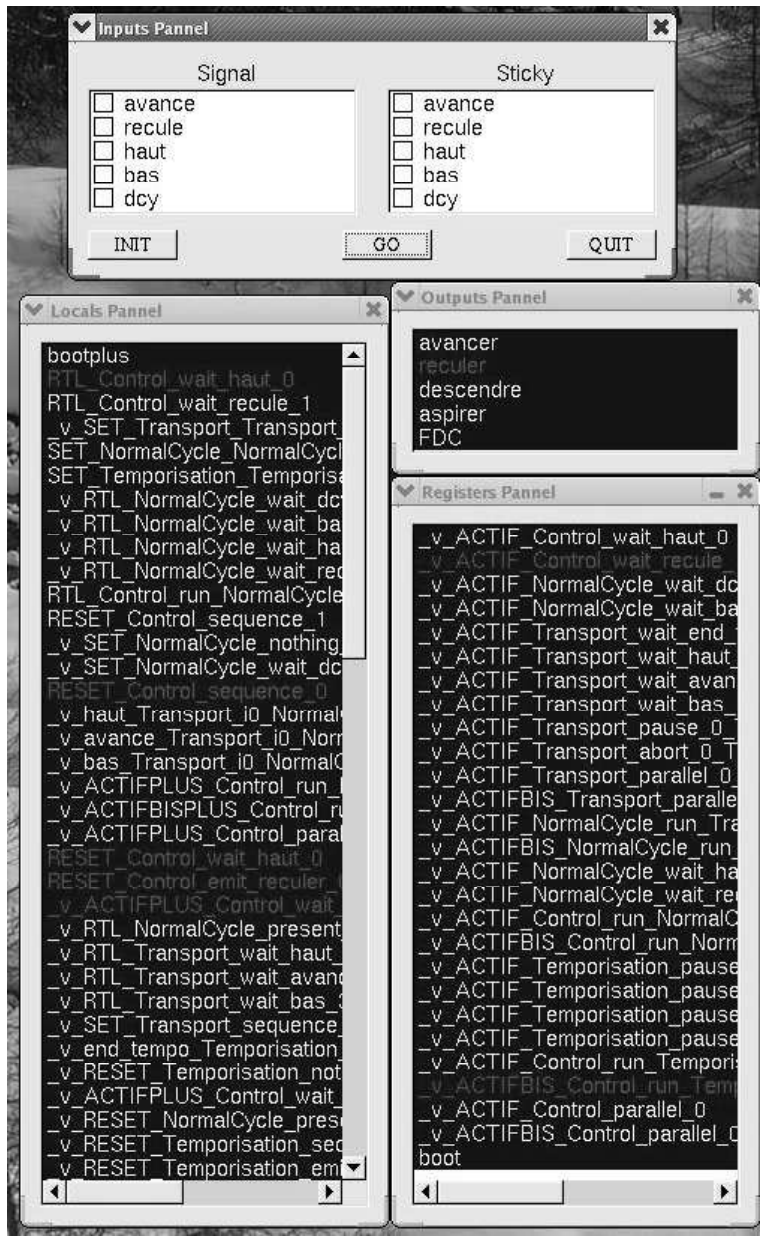}
\caption{Control module simulation panels}
\label{fig:simulation}
\end{figure}

\subsection {Mecatronics System Simulation and Verification}
\label{leexample-verif}

To check the behavior of our implementation with respect to the
specification, we first simulate it and then perform model-checking
verification. Both simulation and verification relies on the
generation of \texttt{blif} format from \texttt{clem} compiler.

Figure~\ref{fig:simulation} shows the result of Control simulation with a graphical tool we have
to simulate blif format modules.

On another hand, to formally prove safety properties we rely on model checking techniques.
In this approach, the correctness of a system with respect to a desired behavior is
verified by checking whether a structure that models the system
satisfies a formula describing that behavior. Such a formula is
usually written by using a temporal logic. Most existing verification
techniques are based on a representation of the concurrent system by
means of a labeled transition system (LTS). 
Synchronous languages are well known to have a clear semantic that allows
to express the set of behaviors of program as LTSs and thus model checking 
techniques are available. Then, they rely on  formal methods to build dependable software.
The same occurs for \lstrl language, the LTS model of a program is naturally encoded in its
equational semantic.

A verification means successfully used for synchronous formalisms
is that of observer monitoring \cite{observer}. According to this technique, a
safety property $\phi$ can be mapped to a program
$\Omega$ which runs in parallel with a program P
and observes its behavior, in the sense that at each instant $\Omega$
reads both inputs and outputs of P. If $\Omega$ detects that P has violated $\phi$
then it broadcasts an "alarm" signal. 
As a consequence, we can rely on model checking based tools to verify property of 
\lstrl language.
But, our approach provides us with separate compilation and requires to be
completed by a modular verification. We aim at proving 
safety properties are preserved through \lstrl language operator application.

To verify that the suction is maintained from the instant
where the cycle begins up to the cycle ends, the following observer can be written in \lstrl.

{\small
\begin{verbatim}

module CheckSuckUp;
Input SuckUp, S;
Output exitERROR;
present SuckUp
        { present S {nothing} else {wait S}}
        else {pause>>emit exitERROR}
end

module SuctionObs:

Input:forward, backward, upward,  downward, 
      StartCycle, Output:MoveFor, MoveBack, 
      MoveDown, SuckUp, EndCycle ;
Output: ERROR;

local exitERROR {
 abort {
  loop {
   present StartCycle {nothing} 
           else {wait StartCycle} >>
   present MoveDown {nothing} 
                    else {wait MoveDown}  >>
   present downward {nothing} 
                    else { wait downward} >>
   present MoveDown {nothing}
           else { present SuckUp 
                     {run CheckSuckUp[upward\S]   >>
                      run CheckSuckUp[MoveFor\S]  >>
                      run CheckSuckUp[forward\S]  >>
                      run CheckSuckUp[MoveDown\S] >>
                      wait downward
                     }
                     else {emit exitERROR}
                }
  }
 when exitERROR >> emit ERROR
 }
}
end
\end{verbatim}
}

To specify the observer we first define a module (CheckSuckUp) which checks wether 
the signal \texttt{SuckUp} is present and  goes in the state where signal \texttt{S}
is present. If \texttt{SuckUp} is absent , exitERROR is emitted.
Calling this module, the observer tests the presence of signal SuckUp  in each possible states  
reached when cylinders move.

To achieve the property checking, we compile a global module made of the Control module in parallel
with the SuctionObs module and we rely on model checker to ensure that ERROR is never emitted.
By the time, we generate the {\sc blif} format back end for the global module and we call
\texttt{xeve} model-checker \cite{xeve} to perform the verification.
In the future, we intend to interface NuSMV \cite{NuSMV} model-checker.

The chosen example is a very simple one  but we hope understandable in the framework of a paper.
Nevertheless, we compiled it globally and in a separate way. The global compilation takes about
2.7 s while the separate one takes 0.6 s on the same machine. We  think that it is a small but
promising result.

\section{Conclusion} 
\label{leconclusion}

In this work, we have presented a new synchronous language \lstrl that supports separate
compilation. We defined its behavioral semantic   giving a meaning to each program
and allowing us to rely on formal methods to achieve verification.
Then, we also defined an equational semantic to get a means to really  compile
programs in a separate way.
Actually, we have implemented the \texttt{clem/clef} compiler.
This compiler is the core of the design chain (see section \ref{compilation-scheme}) we have to
specify control-dominated process from different front-ends: a graphical editor devoted
to automata drawing, or direct \lstrl language specification to several families of
back-ends:
\begin{itemize}
\item \textit{code generation}: we generate either executable code as C code or
model-driven code: Esterel, Lustre code for software applications and Vhdl for
harware targets.
\item \textit{simulation tools}: thanks to the \textit{blif} format generation we can rely
our own simulator (\textit{blif\_simul}) to simulate \lstrl programs.
\item \textit{verification tools}: {\sc blif} is a well-suited format to 
several model-checkers(xeve, sis) and has its automata equivalence verifier 
(blif2autom, blifequiv).
\end{itemize}

In the future, we will focus on  three main directions.
The first one concerns our compilation methodology. Relying on an equational semantic to
get modular compilation could lead to generate inefficient code.
To avoid this drawback, we plan to study others equational semantic rules (in particular
for  parallel and run statements) more suited for optimization.
The second improvement we aim at, is the extension of the language. To be able to deal with
control-dominated systems with data (like sensor handling), we will extend the syntax
of the language on the first hand. On the other hand,
we plan to integrate abstract interpretation techniques (like
polyhedra intersection, among others) \cite{Cousot} to
take into account data constraints in control. Moreover, we also need to communicate with
signal processing or automation world through their specific tool Matlab/Simulink
(http://www.mathworks.com). 
Another language extension is to allow a bound number of parallel operators. This extension
is frequently required by users to specify their applications. Semantic rules for
this new bound parallel operator cannot be straightly deduced from the actual rules we have,
and require a deep change but then  would improve  \lstrl expressiveness.
Finally, we are interested in improving our verification means.
The synchronous approach provides us with well-suited models 
to apply model checking techniques to \lstrl programs.
The more efficient way seems to directly interface a powerful model-ckecker (as NuSMV \cite{NuSMV})
and to be able to run its property violation scenarios in our simulation tool.
Moreover, our modular approach opens new ways to modular verification. We need to
prove that \lstrl operators preserve properties: if a program $P$ verify a property $\phi$,
then all program using $P$  should verify a property $\phi'$ such that 
the ``restriction'' of $\phi'$ to $P$ implies $\phi$.

\bibliographystyle{plain}
\bibliography{biblio}

\begin{thebibliography}{10}

\bibitem{SyncChart}
C.~André, H.~Boufa\"{i}ed, and S.~Dissoubray.
\newblock Synccharts: un modèle graphique synchrone pour système réactifs
  complexes.
\newblock In {\em Real-Time Systems(RTS'98)}, pages 175--196, Paris, France,
  January 1998. Teknea.

\bibitem{berry-book}
G.~Berry.
\newblock {\em {The Constructive Semantics of Pure Esterel}}.
\newblock Draft Book, available at: http://www.esterel-technologies.com 1996.

\bibitem{berry}
G.~Berry.
\newblock {The Foundations of Esterel}.
\newblock In G.~Plotkin, C.~Stearling, and M.~Tofte, editors, {\em {Proof,
  Language, and Interaction, Essays in Honor of Robin Milner}}. MIT Press,
  2000.

\bibitem{xeve}
Amar Bouali.
\newblock Xeve , an esterel verification environment.
\newblock Technical report, CMA-Ecole des Mines, 1996.

\bibitem{NuSMV}
A.~Cimatti, E.~Clarke, E.~Giunchiglia, F.~Giunchiglia, M.~Pistore, M.~Roveri,
  R.~Sebastiani, and A.~Tacchella.
\newblock {NuSMV~2: an OpenSource Tool for Symbolic Model Checking}.
\newblock In Ed~Brinksma and Kim~Guldstrand Larsen, editors, {\em Proceeeding
  CAV}, number 2404 in LNCS, pages 359--364, Copenhagen, Danmark, July 2002.
  Springer-Verlag.

\bibitem{Cousot}
P.~Cousot and R.~Cousot.
\newblock {On Abstraction in Software Verification}.
\newblock In Ed~Brinksma and Kim~Guldstrand Larsen, editors, {\em Proceeeding
  CAV}, number 2404 in LNCS, pages 37,56, Copenhagen, Danmark, July 2002.
  Springer-Verlag.

\bibitem{cec}
S.A. Edwards.
\newblock Compiling esterel into sequential code.
\newblock In {\em Proceedings of the 7th International Workshop on
  Hardware/Software Codesign (CODES 99)}, pages 147--151, Rome, Italy, May
  1999.

\bibitem{denotational}
M.~Gordon.
\newblock {\em The Denotational Description of Programming Languages}.
\newblock Springer-Verlag, 1979.

\bibitem{HAL93}
N.~Halbwachs.
\newblock {\em Synchronous Programming of Reactive Systems}.
\newblock Kluwer Academic, 1993.

\bibitem{observer}
N.~Halbwachs, F.~Lagnier, and P.~Raymond.
\newblock Synchronous observers and the verification of reactive systems.
\newblock In M.~Nivat, C.~Rattray, T.~Rus, and G.~Scollo, editors, {\em Third
  Int. Conf. on Algebraic Methodology and Software Technology, AMAST'93},
  Twente, June 1993. Workshops in Computing, Springer Verlag.

\bibitem{HP85}
D.~Harel and A.~Pnueli.
\newblock On the development of reactive systems.
\newblock In {\em NATO, Advanced Study institute on Logics and Models for
  Verification and Specification of Concurrent Systems}. Springer Verlag, 1985.

\bibitem{causality-modularity}
C.~Huizing and R.~Gerth.
\newblock Semantics of reactive systems in abstract time.
\newblock In {\em Real Time: Theory in Practice, Proc of REX workshop}, pages
  291--314. W.P. de Roever and G. Rozenberg Eds,LNCS, June 1991.

\bibitem{jacky}
D.~Potop-Butucaru and R.~De Simone.
\newblock {\em Formal Methods and Models for System Design}, chapter
  Optimizations for Faster Execution of Esterel Programs.
\newblock Gupta, P. LeGuernic, S. Shukla, and J.-P. Talpin, Eds ,Kluwer, 2004.

\bibitem{Shiple}
T.~Shiple.
\newblock {\em Formal Analysis of Cyclic Circuits}.
\newblock PhD thesis, University of California, 1996.

\bibitem{esterelstudio}
Esterel Technologies.
\newblock Esterel studio suite, www.estereltechnologies.com.

\bibitem{saxo-rt}
D.~Weil, V.~Bertin, E.~Closse, M.~Poize, P.~Venier, and J.~Pulou.
\newblock Efficient compilation of esterel for real-time embedded systems.
\newblock In {\em Proceedings of the 2000 International Conference on
  Compilers, Architecture, and Synthesis for Embedded Systems,}, pages 2--8,
  San Jose, California, United States, November 2000.

\end{thebibliography}

\newpage

\appendix
\section{LE Grammar}
\label{legrammar}

In this appendix, we describe the complete grammar of the \lstrl language. That description
supports the following agreements:
\begin{itemize}
\item <> notation represents tokens, for instance <module> represents the name \textit{module};
\item two specific tokens are introduced: IDENT for identifier and STRING to denote a usual
string;
\item the notation $\star$ and $+$ are used for repetition: $signal\_name\star$ means a  
number of $signal\_name$, possibly 0,while  $signal\_name+$ means at least one occurrence;
\item the single charater are straighly written (as \{,\},[,], and $\backslash$).
\item the character \# denotes the empty word;
\end{itemize}

{\small

\begin{verbatim}

program: <module> module_name ':' module_interface module_body <end>;
         
module_interface : input_signal_list output_signal_list run_decl_list;

input_signal_list : # | <Input:> signal_name+ ';' ;
output_signal_list : #  | <Output:> signal_name+ ';' ;
run_decl_list : #  | <Run:> run_declaration+ ;
run_declaration: path ':' module_name;
        
module_body : instruction | automaton ;

instruction : statement | '{' instruction '}' ;

statement : parallel
          | sequence 
          | present
          | loop  
          | wait
          | emit
          | abort  
          | nothing 
          | pause 
          | halt 
          | local
          | run 
          ;

parallel : instruction '||' instruction ;
sequence :  instruction '>>' instruction ;
present : <present> xi_expression  instruction   <else> instruction ;
loop :  <loop> '{' instruction '}' ;
wait :   <waitL>  signal_name ;
emit : <emit> signal_name ;
abort : <abort> '{' instruction '}' <when> signal_name ;
pause : <pause> ;
nothing : <nothing> ;
halt : <halt> ;
local :  <local> signal_name+ '{' instruction '}'
run : <run> module_name  renaming ;
renaming : #  | '[' single_renaming+ ']' ;
single_renaming :  signal_name '\' signal_name  


automaton : <automaton> state+ transition_def ; 
state : <state> state_name opt_final opt_run  action ';' ;
opt_run: #  | run ;
transition_def :  <transition> transition+ ; 
transition : opt_initial opt_final  opt_source_state  trigger  action opt_target_state ;
opt_source_state: # | state_name;
opt_target_state: # | '->' state_name;
opt_initial : # | <initial> ;
opt_final:  # | <initial> ;
trigger:  # | xi_expression ;
action: # | '/' signal_name+ ;
xi_expression : xi_expression <or> xi_expression
              | xi_expression <and> xi_expression
              | <not> xi_expression
              | '{' xi_expression '}'
              | signal_name
              ;

signal_name : IDENT;
module_name : IDENT;
path : STRING;

\end{verbatim}
}

\newpage

\section{PERT Algorithms}

\subsection{First Step of PERT ALGORITHM}
\label{pertalgorithm_1}

The following algorithm is the first step of the overall PERT algorithm we implement.
It builds a forest of variable dependency trees.

{\small
\begin{verbatim}
for each equation xi=fi(...,xj,...)
begin
   for all j needed by fi
   begin
      Upstream[i].add(j);
      Downstream[j].add(i)
   end
end
\end{verbatim}
}

\subsection{Second Step of PERT Algorithm}
\label{pertalgorithm_2}

The second step of the PERT algorithm we implement consists in the propagation of
the \textsl{Early} Dates from the inputs and the registers, to the outputs. Similarly,
the \textsl{Late} Dates are propagated from the outputs to the inputs and the
registers according to the following algorithm:

{\small
\begin{verbatim}
for each variable id i
begin
   if(Upstream[i] = empty set)
   begin
      /* final output */
      late[i]=0
      for each j in Downstream[i] 
      begin
         late_propagation(j,1)
      end 
   end
   if(Downstream[i] = empty set)
   begin
      /* real input or constante */
      early[i]=0
      for each j in Upstream[i] 
      begin
         early_propagation(j,1)
      end 
   end
end

function late_propagation(id,date)
begin
  if(late[id] < date)
  begin
    late[id]=date
    for each j in Dowstream[id] 
    begin
       late_propagation(j,date+1)
    end
  end
end

function early_propagation(id,date)
begin
  if(early[id] < date)
  begin
    early[id]=date
    for each j in Upstream[id] 
    begin
       early_propagation(j,date+1)
    end
  end
end
\end{verbatim}
}

\newpage
\section{LE Control Example Code}

\label{lecontrolcode}

In this appendix, we detail the \lstrl code for the \textit{Control} example described
in section \ref{leexample}.

\subsection{Control Module Specification}

The main file of the \textit{Control} example is  \textit{Control.le}.
We give its content:

{\small 
\begin{verbatim}

;;====================================================
;; LE specification for a mecatronic system
;; Main file: Control specification
;;====================================================


module Control:

Input:  forward, backward, upward, downward, StartCycle;
Output: MoveFor, MoveBack, MoveDown, SuckUp, EndCycle;

Run: "/home/ar/GnuStrl/work-ar/TEST/control/" : Temporisation;
     "/home/ar/GnuStrl/work-ar/TEST/control/" : NormalCycle;

local start_tempo, end_tempo {

     { wait upward >> emit MoveBack >> wait backward >> run NormalCycle}
  ||
     { run Temporisation}
}

end
\end{verbatim}
}
 
The \textit{Control} module calls two external modules
\textit{Temporisation} and \textit{NormalCycle}. The paths to
\textit{Temporisation.le} and \textit{NormalCycle.le} files where the
respective \lstrl codes of these called modules are, is given in
\textit{Control} module interface.  During compilation, a file
\textit{temporisation.lec} (resp \textit{NormalCyle.lec}) is searched
in the compilation library. If \textit{Temporisation} (resp
\textit{NormalCycle}) has not been already compiled then it is
compiled. Thus, in both cases, the compiled code is included in
\textit{Control} module code.

\subsection{Temporisation module Specification}

{ \small
\begin{verbatim}

;;====================================================
;; LE specification for a mecatronic system
;; Temporisation  specification
;;====================================================

module Temporisation :

Input:  start_tempo;
Output: end_tempo;

present start_tempo {
 pause >> pause >> pause >> pause >> emit end_tempo }
else nothing

end
\end{verbatim}
}

\subsection{NormalCycle module Specification}

{ \small

\begin{verbatim}

;;====================================================
;; LE specification for a mecatronic system
;; Normal cycle  specification
;;====================================================

module Transport :

Input: end_tempo, upward, forward, downward;
Output: MoveDown,MoveFor, SuckUp;

local exitTransport {

    { emit MoveDown >> wait end_tempo >> wait upward >> emit MoveFor 
      >> wait forward >> emit MoveDown >> wait downward  >> emit exitTransport
    }
 ||

    abort
      { loop { pause >> emit SuckUp }}
    when exitTransport
}

end 

module NormalCycle :

Input: StartCycle, downward, upward, backward, end_tempo, forward;
Output: start_tempo, MoveDown,MoveBack, EndCycle, MoveFor, SuckUp;

{ present StartCycle { nothing} else wait StartCycle }

>>

{
  loop { emit MoveDown >> wait downward >> emit start_tempo  >> run Transport 
         >> wait upward >> emit MoveBack 
         >> wait backward >> emit EndCycle }
} 

end
\end{verbatim}
}

The \textit{NormalCycle} module called itself a \textit{Transport} module, but contrary to
\textit{Control} module, the specification of the called module is
given  in the same file. Thus, no  path has to be supplied in \textit{NormalCycle} interface.

\newpage

\section{Condition Law Expansion}
\label{appendix-lecondlaw}

In this appendix, we discuss how a term from $\xi$ algebra resulting of the application
of the condition law is expanded in a pair of boolean values in $\Bool$..
Let us consider a $\xi$ term $X$. We recall that $X$ is isomorphic to a pair of boolean
($X_{def}$, $X_{va;}$) (see section \ref{lecondlaw}) and we want to prove the following equalities:
$(X \blacktriangleleft c)_{def}  =  X_{def}.c$ and
$(X \blacktriangleleft c)_{val}  =  X_{val}.c$, where $c \in \Bool$.

These equalities are very useful for implementing the condition law in the compilation phasis.

First, relying on the definition of the isomorphism between $\xi$ algebra and 
$\Bool \times \Bool$, we can expand the encoding of the condition law as follow:

\begin{center}
\begin{tabular}{|c | c | c | c || c|  c | c |}
\hline
$X$ & {$X_{def}$} & {$X_{val}$} & $c$ & $(X \blacktriangleleft c)$ & 
                                  $(X \blacktriangleleft c)_{def}$ & 
                                  $(X \blacktriangleleft c)_{val}$ \\
\hline
1 & 1 & 1 & 0 & $\bot$ & 0 & 0 \\
0 & 1 & 0 & 0 & $\bot$ & 0 & 0 \\
$\top$ & 0 & 1 & 0 & $\bot$  & 0 & 0 \\
$\bot$ & 0 & 0 & 0 & $\bot$  & 0 & 0\\
\hline
1 & 1 & 1 & 1 & 1 & 1 & 1 \\
0 & 1 & 0 & 1 & 0 & 1 & 0 \\
$\top$ & 0 & 1 & 1 &$\top$ &  0 & 1 \\
$\bot$ & 0 & 0 & 1 & $\bot$ & 0 & 0\\
\hline
\end{tabular}
\end{center}

where $c \in \Bool$.

Thus, we can deduce:

\begin{tabular}{l l}
$(X \blacktriangleleft c)_{def}$ & = $X_{def}.X_{val}.c\ +\ X_{def}.\overline{X_{val}}.c $\\
& $ = X_{def}.c.(X_{val} + \overline{X_{val}})$\\
& $ = X_{def}.c$\\
\\
$(X \blacktriangleleft c)_{val}$ & = $X_{def}.X_{val}.c\ +\ \overline{X_{def}}.X_{val}.c $\\
& $ = X_{val}.c.(X_{def} + \overline{X_{def}})$\\
& $ = X_{val}.c$\\
\end{tabular}

\newpage

\section{LE Statement Circuit Description}
\label{lecircuit}

In this appendix, we show the circuits corresponding to \lstrl statement.
We rely on them to compute the equational semantic of each \lstrl operator.

\begin{figure}[htbp]
  \centering
      \subfigure[Circuit for \fixed{nothing}]{
       \includegraphics[width=6cm,ext=.eps]{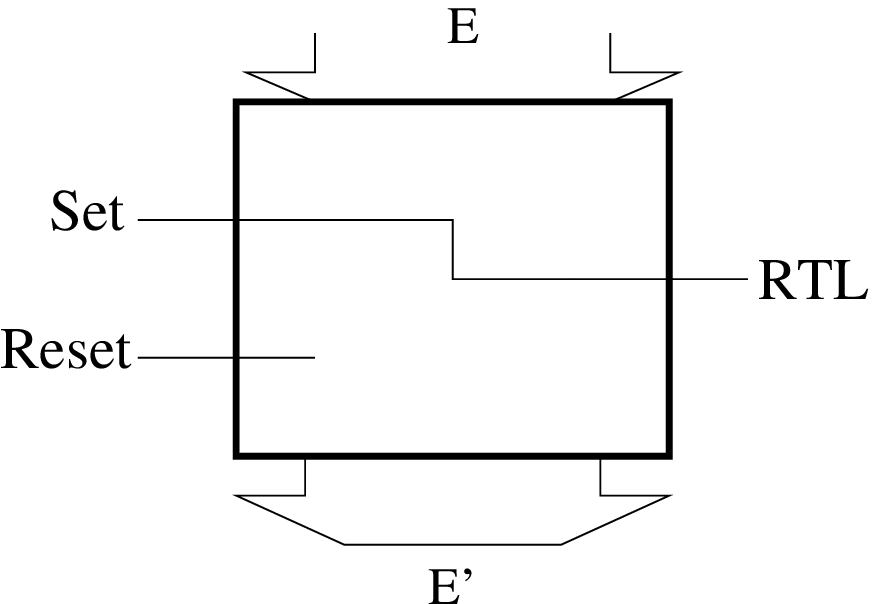}
        \label{nothing}
      } \hfill
      \subfigure[Circuit for \fixed{halt}]{
        \includegraphics[width=6cm,ext=.eps]{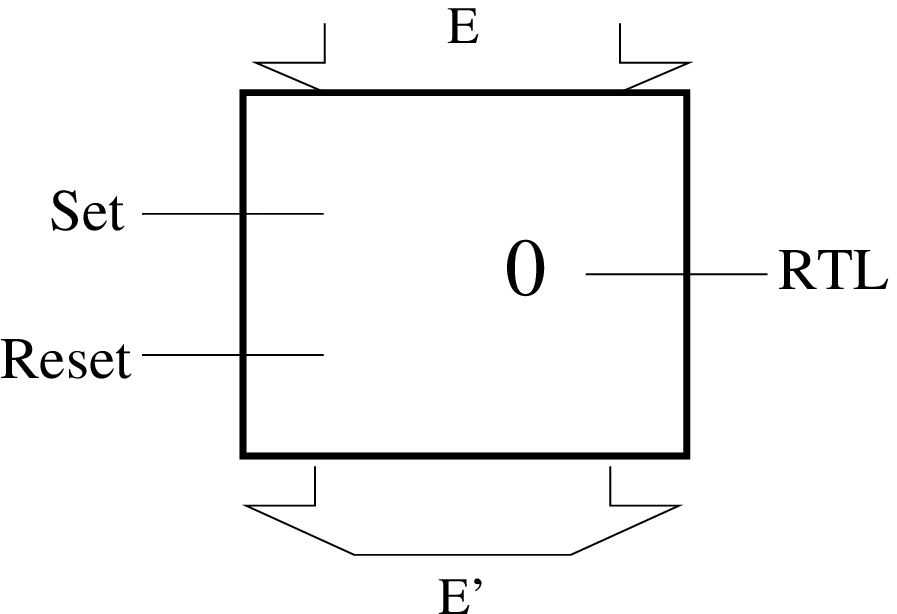}
        \label{halt}
      }
    \caption{Basic \lstrl statements circuit semantic}
     \label{nothing-halt}
\end{figure}

\begin{figure}[htbp]
\centerline{\epsfxsize=6cm \epsfbox{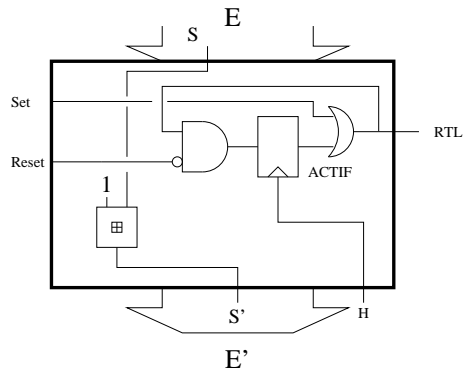}}
\caption{Circuit for \fixed{emit}$S$}
\label{emit}
\end{figure}

\begin{figure}[htbp]
  \centering
      \subfigure[Circuit for \fixed{pause}]{
        \includegraphics[width=6cm,ext=.eps]{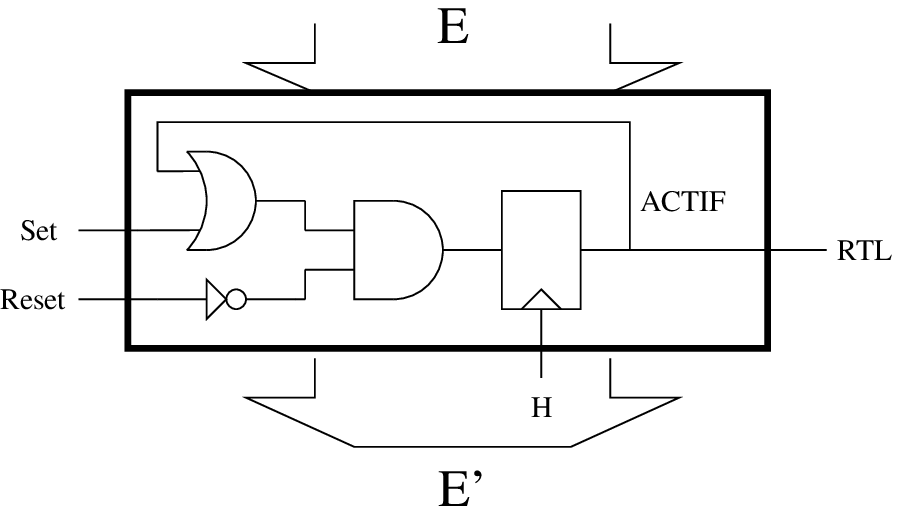}
        \label{pause}
      } \hfill
      \subfigure[Circuit for \fixed{wait}]{
        \includegraphics[width=6cm,ext=.eps]{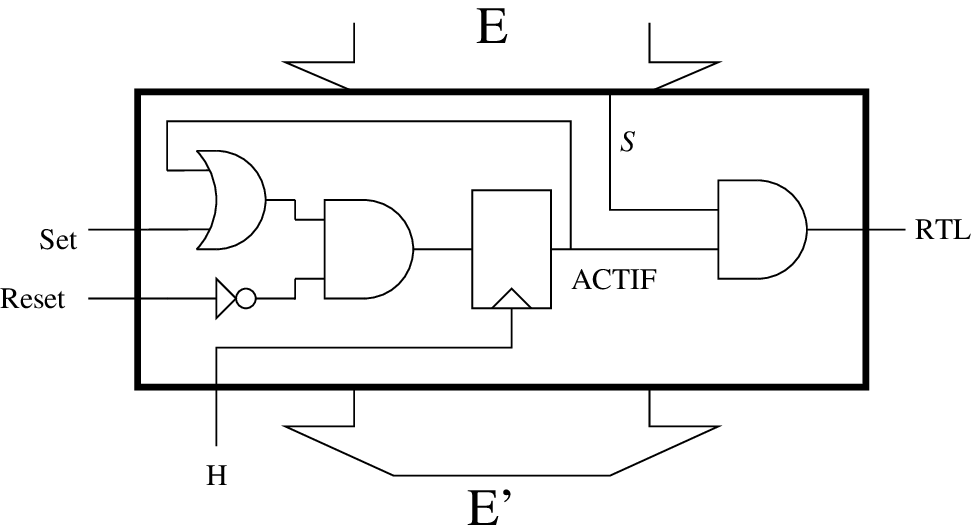}
        \label{wait}
      }
    \caption{Pause and Wait \lstrl statements circuit semantic}
     \label{pause-wait}
\end{figure}

\begin{figure}[htbp]
\centerline{\epsfxsize=10cm \epsfbox{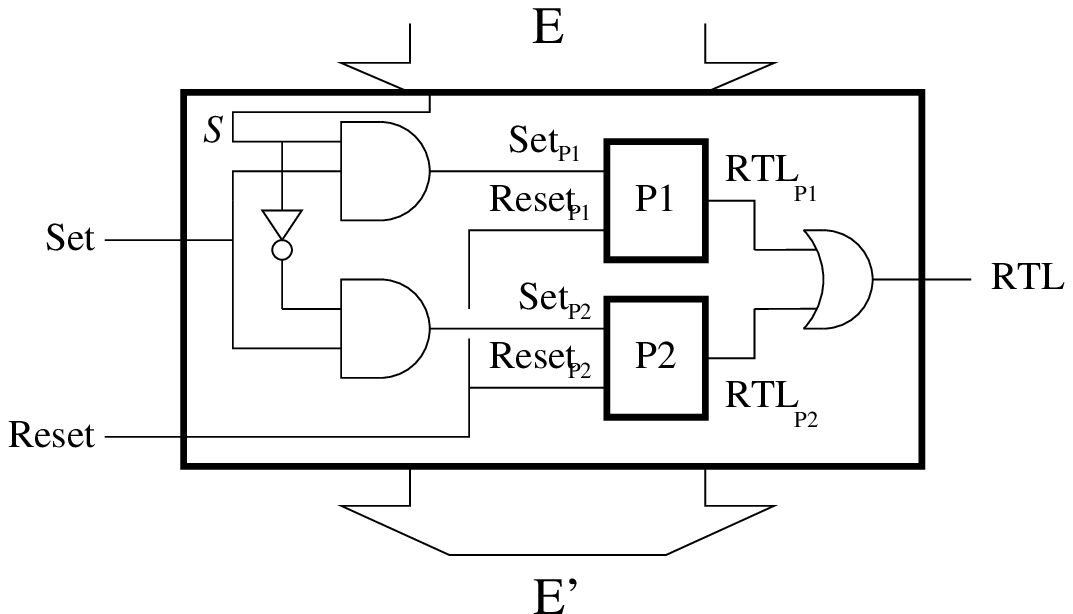}}
\caption{Circuit for $\fixed{Present}\ S \{{\rm P}_1\} \fixed{else} \{{\rm P}_2\}$}
\label{present}
\end{figure}

\begin{figure}
\centerline{\epsfxsize=8cm \epsfbox{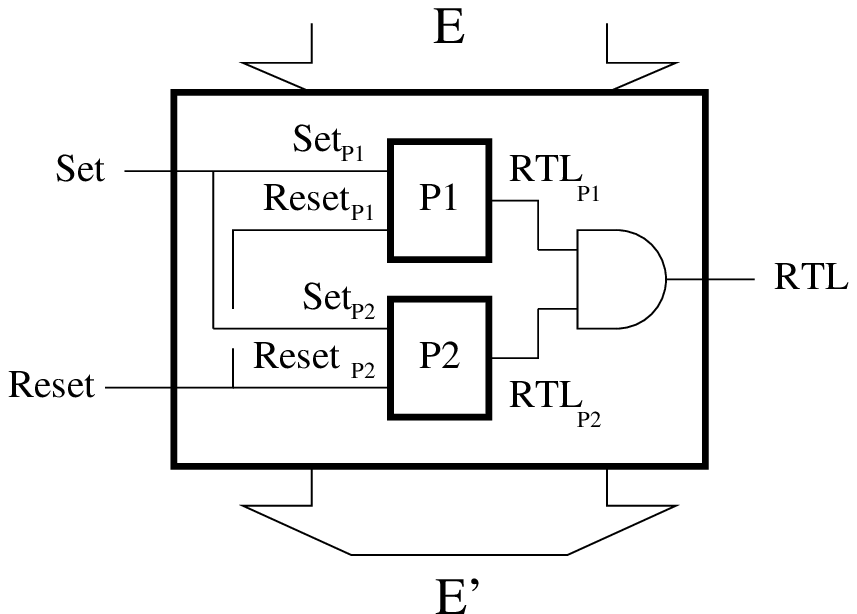}}
\caption{Circuit for${\rm P}_1 \| {\rm P}_2$}
\label{parallel}
\end{figure}

\begin{figure}
\centerline{\epsfxsize=9cm \epsfbox{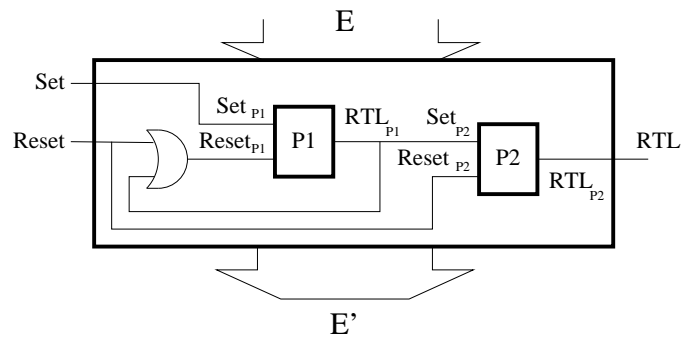}}
\caption{Circuit for ${\rm P}_1 \gg {\rm P}_2$}
\label{sequence}
\end{figure}

\begin{figure}
\centerline{\epsfxsize=10cm \epsfbox{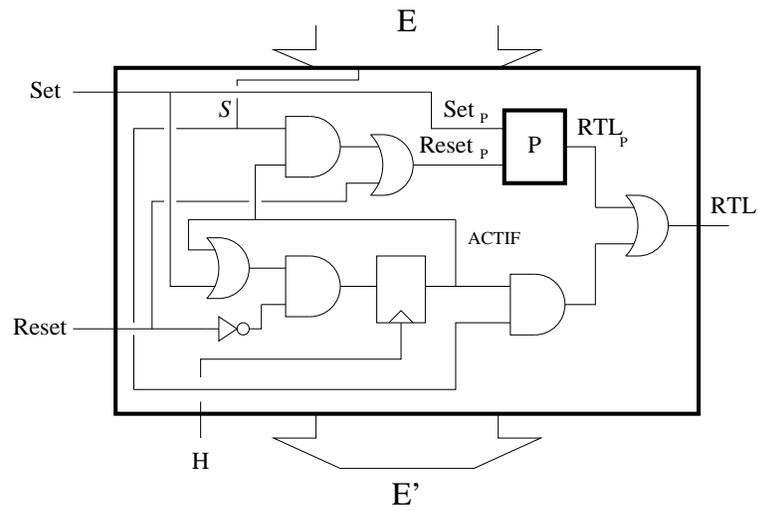}}
\caption{Circuit for abort P when $S$}
\label{abort}
\end{figure}

\begin{figure}
\centerline{\epsfxsize=10cm \epsfbox{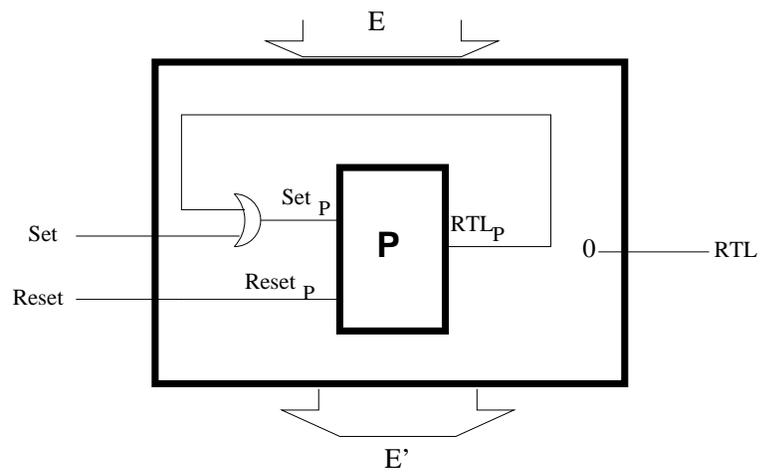}}
\caption{Circuit for loop \{P\}}
\label{loop}
\end{figure}

\begin{figure}
\centerline{\epsfxsize=10cm \epsfbox{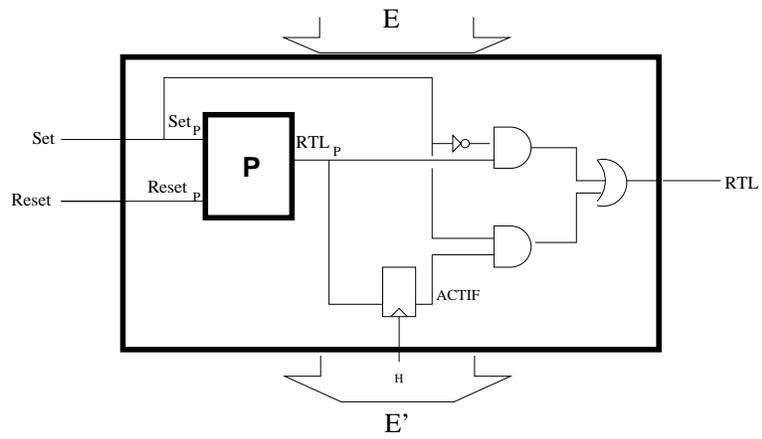}}
\caption{Circuit for \fixed{run}\{P\}}
\label{run}
\end{figure}

\begin{figure}
\centerline{\epsfysize=9cm \epsfbox{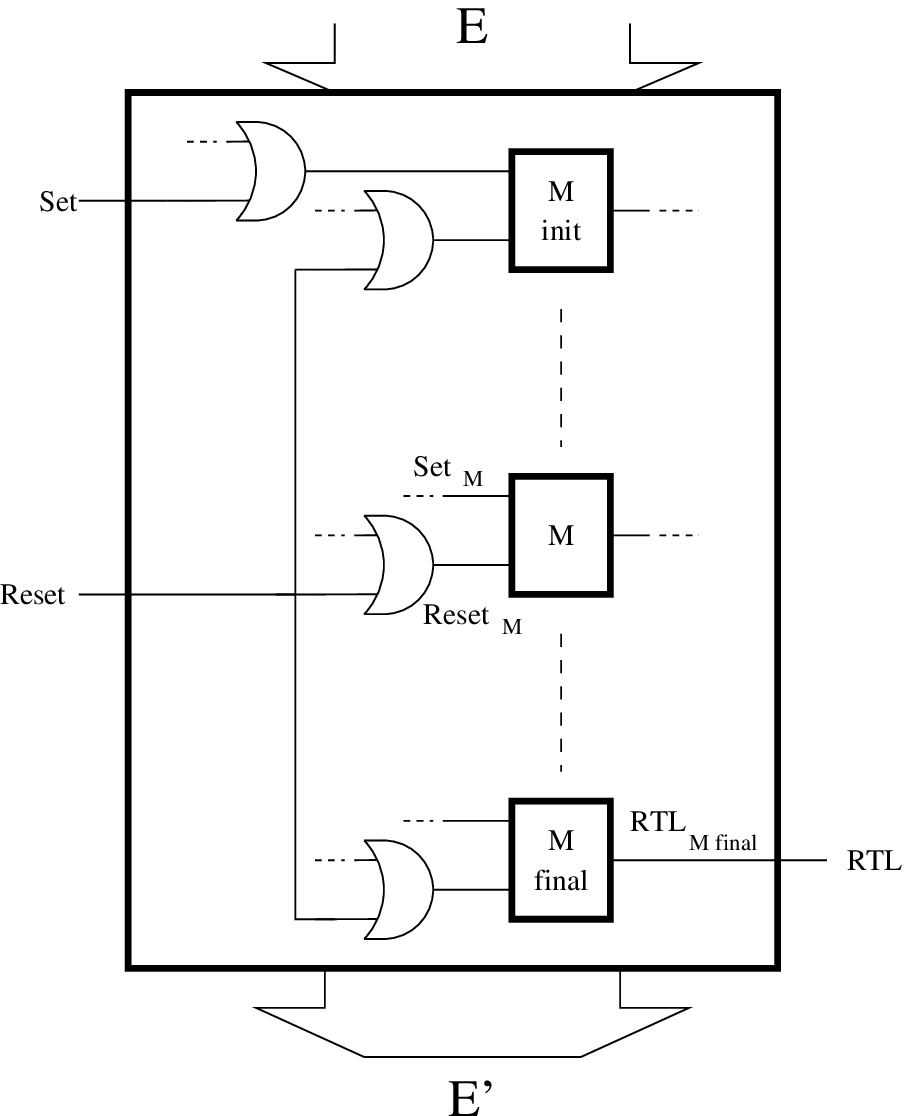}}
\caption{Circuit for ${\cal A}({\rm M}_{\rm init}, ..,{\rm M},..., {\rm M}_{\rm final})$}
\label{autom}
\end{figure}

\end{document}